%% file: main.tex
\documentclass{llncs}

\usepackage{amsmath,amssymb,stmaryrd,latexsym,mathtools}
\usepackage[bbgreekl]{mathbbol}
\usepackage{graphicx}
\usepackage{proof}
\usepackage{url}
\usepackage[T1]{fontenc}
\usepackage{times,textcomp}
\usepackage{picinpar}
\usepackage{ifthen}
\usepackage{enumerate}
\usepackage{color}

\newboolean{LONG}
\setboolean{LONG}{true}

\ifLONG
\else
\fi

\input{macros}

\begin{document}

\title{Program Equivalence in Linear Contexts}
\author{Yuxin Deng\inst{1,2} \and Yu Zhang\inst{2}}
\institute{
Department of Computer Science and Engineering,
Shanghai Jiao Tong University, China
\and 
State Key Laboratory of Computer Science, \\ 
Institute of Software, Chinese Academy of Sciences, Beijing, China
}

\maketitle

\input{intro}

\input{lpcf}

\input{nlpcf}

\input{conclu}


\end{document}

%% file: macros.tex
\newcommand{\defeq}{\stackrel{\mathrm{def}}{=}}

\newcommand{\allconverge}{\downdownarrows}
\newcommand{\alldiverge}{\upuparrows}

\newcommand{\bang}{\,!}

\newcommand{\cbind}[3]{\mathtt{bind} \ {#1} = {#2} \ \mathtt{in} \ {#3}}

\newcommand{\cpred}{\mathtt{pred}}
\newcommand{\csucc}{\mathtt{succ}}
\newcommand{\czero}{\mathtt{iszero}}

\newcommand{\cif}[3]{\mathtt{if} \; {#1} \; \mathtt{then} \; {#2} \; \mathtt{else} \; {#3}}
\newcommand{\ctrue}{\mathtt{true}}
\newcommand{\cfalse}{\mathtt{false}}
\newcommand{\cfix}{\mathtt{fix}}
\newcommand{\clet}[3]{\mathtt{let} \ {#1} = {#2} \ \mathtt{in} \ {#3}}

\newcommand{\cpair}[1]{\langle{#1}\rangle}
\newcommand{\cproj}{\mathtt{proj}}

\newcommand{\cval}{\mathtt{val}}
\newcommand{\choice}{\sqcap\,}

\newcommand{\context}{\mathcal{C}}

\newcommand{\converge}{\Downarrow}

\newcommand{\diverge}{\Uparrow}

\newcommand{\evcontext}{\mathcal{E}}

\newcommand{\FV}{\mathit{FV}}
\newcommand{\FLV}{\mathit{FLV}}
\newcommand{\FNV}{\mathit{FNV}}

\newcommand{\hole}{[\,]}

\newcommand{\infini}{\pmb{\Omega}}

\newcommand{\intransto}[1]{\xhookrightarrow{\;{#1}\;}}

\newcommand{\ldot}{\, . \, }

\newcommand{\lintransto}[1]{\,\circ\hspace{-0.5em}\xrightarrow{\,{#1}\,}}
\newcommand{\linto}{\multimap}

\newcommand{\nulltrace}{\epsilon}

\newcommand{\preord}{\sqsubseteq}

\newcommand{\product}{\,\&\,}

\newcommand{\prog}{\mathcal{P}\!\mathit{rog}}

\newcommand{\reduce}{\leadsto}

\newcommand{\Reduceto}{\Downarrow}
\newcommand{\set}[1]{\{{#1}\}}

\newcommand{\tbool}{\mathsf{Bool}}
\newcommand{\tensor}{\otimes}
\newcommand{\tnat}{\mathsf{Nat}}
\newcommand{\tcomp}{\mathsf{T}}

\newcommand{\transto}[1]{\xrightarrow{\,{#1}\,}}
\newcommand{\trace}{\mathcal{T}\!\mathit{r}}

%% file: intro.tex
\begin{abstract}
Program equivalence in linear contexts, where programs are used or executed {\em exactly once}, 
is an important issue in programming languages.
However, existing techniques like those based on bisimulations and logical relations only target 
at contextual equivalence in the usual (non-linear) functional languages, 
and fail in capturing non-trivial equivalent programs in linear contexts, 
particularly when non-determinism is present. 

We propose the notion of {\em linear contextual equivalence} to formally characterize such 
program equivalence, as well as a novel and general approach to studying it in higher-order 
languages, based on labeled transition systems specifically designed for functional languages. 
We show that linear contextual equivalence indeed coincides with trace equivalence.
We illustrate our technique in both deterministic (a linear version of PCF) and 
non-deterministic (linear PCF in Moggi's framework) functional languages.
\end{abstract}

\section{Introduction}
\label{sec:intro}

Contextual equivalence is an important concept in programming languages and can be used to 
formalize and reason about many interesting properties of computing systems. 
For functional languages, there are many techniques that can help to prove contextual equivalence.
Among others, applicative bisimulations~\cite{Abr90,How96} 
and logical relations~\cite{Plo80,Sta85} are particularly successful.
\ifLONG

\else
\fi
On the other side, linear logic (and its term correspondence often known as linear $\lambda$-calculus) 
has seen significant applications in computer science ever since its birth, 
due to its native mechanism of describing restricted use of resources. 
For example, the linear $\lambda$-calculus provides the core of a functional programming language 
with an expressive type system, in which statements like ``this resource will be used exactly once'' 
can be formally expressed and checked. Such properties become useful when introducing imperative 
concepts into functional programming~\cite{Hof00}, structural complexity theory \cite{Hof03}, 
or analyzing memory allocation \cite{WW01}. 
\ifLONG
Moreover, linear $\lambda$-calculus, when equipped with dependent types, can serve as a representation 
language within a logical framework, a general meta-language for the formalization 
of deductive systems \cite{CP96}.
\else
\fi

Introducing linearity also leads to novel observation over program equivalences. 
In particular, if we consider a special sort of contexts where candidate programs must be used linearly 
(we call these contexts \emph{linear contexts}), program equivalence with respect to these contexts 
should be a coarser relation than the usual notion of contextual equivalence, 
especially when non-determinism is present. 
For instance, take Moggi's language for non-determinism~\cite{Mog91}, where we have a primitive 
$\choice$ for non-deterministic choice (same as the internal choice in
CSP \cite{Hoa85}), 
and consider the following two functions:
\[
f_1 \defeq \cval(\lambda x \ldot \cval(0) \choice \cval(1)) ,
\qquad
f_2 \defeq \cval(\lambda x \ldot \cval(0)) \choice \cval(\lambda x \ldot \cval(1)).
\]
Existing techniques such as bisimulation or logical relations 
distinguish these two functions. 
In fact, it is easy to show that 
they are not equivalent in arbitrary contexts, 
by considering, e.g., the context 
\[
\cbind{f}{[\_]}{} \cbind{x}{f (0)}{}\cbind{y}{f (0)}{\cval(x = y)} .
\]
The context makes a double evaluation of the function by applying it to concrete arguments
(noticing that Moggi's language enforces a call-by-value evaluation of non-deterministic computations): 
with the first function $f_1$, the two evaluation of $f(0)$ can return different values since 
the non-deterministic choice is inside the function body; 
with the second function $f_2$, the non-deterministic choice is made before both evaluations 
of $f(0)$ and computation inside the function is deterministic, so the two evaluations always 
return the same value. 
But if we consider only linear contexts, where programs will be evaluated {\em exactly once}, 
then the two functions must be equivalent.
However, no existing technique, at least to the best of our knowledge, can characterize such an equivalence 
relation with respect to linear contexts. 

\ifLONG
\subsection{Related work}
\else
\vspace*{-1.2em}
\subsubsection{Related work.}
\fi
The motivation of the work first comes from the second author's work on building a logic (namely {CSLR}) 
for reasoning about {\em computational indistinguishability},
which is an essential concept in complexity-theoretic cryptography and helps to define 
many important security criteria~\cite{Zha10,Gol01}. 
The CSLR logic is based on a functional langage which characterizes probabilistic polynomial-time 
computations by typing, where linearity plays an important role. 
A rule that can identify program equivalence in linear contexts\footnote{
More precisely, in the setting of cryptography we consider adversaries that can call a procedure for 
{\em polynomial} number of times. It has been proved, with certain constraints,  that such adversaries 
cannot achieve more than those who call the program {\em only once}, 
which can be seen as a linear context in CSLR.}
can help to simplify many proofs, e.g., the IND-CPA proof of the El-Gamal encryption, 
which is currently in the form of so-called game-based proofs~\cite{NZ10}. 
Although the language of the CSLR logic is probabilistic, a general proof technique of linear 
contextual equivalence is missing from the literature, particularly in the setting of purely 
non-determinism where there exist programs that are equivalent in linear contexts but not in general, 
as we described previously. 

Program equivalence with respect to \emph{non-linear} contexts has been widely investigated. 
Logical relations are one of the powerful tools for proving contextual equivalence in typed 
lambda-calculi, 
in both
operational~\cite{Pit00,Pit05,BPR00} and denotational settings~\cite{Plo80,Mit96,GLLN08}. 
They are defined by induction on types, hence are relatively easy to use. 
But it is known that completeness of (strict) logical relations are often hard to achieve, 
especially for higher-order types. It is even worse for monadic types, particularly 
when non-determinism is present~\cite{LNZ06}.

Characterization in terms of simulation relation has been studied in functional 
languages~\cite{How96,Gor95,Pit97,Jef99}, as well as languages with linear type systems~\cite{Bie00}. 
Due to the higher-order features of the languages, it is difficult to directly prove 
the precongruence property of similarity.
A common feature crucial to this line of research is then to follow Howe's approach~\cite{How96},
which requires to first define a precongruence candidate, a precongruence relation 
by construction, and then to show the coincidence of that relation with simulation. 
An alternative approach, such as environmental
bisimulation proposed in \cite{SKS07}, has a built-in congruence
property, but then the definition itself has very complex conditions. 

\ifLONG
\subsection{Contribution}
\else 
\vspace*{-1.2em}
\subsubsection{Contribution.}
\fi
In this paper we consider contextual equivalence with respect to \emph{linear} contexts only. 
Our approach is developed in a linear version of PCF and we propose a formal definition of 
the so-called {\em linear contextual equivalence}, which characterizes the notion of program 
equivalence when they are used only once. 
We give a sound and complete characterization of the linear contextual equivalence in terms 
of \emph{trace equivalence}, based on appropriate labeled transition semantics for terms.
\ifLONG
In order to show the congruence property of trace equivalence, we exploit the internal structure 
of linear contexts, instead of relying on Howe's approach. 
\else 
\fi

While term transitions are a relatively standard concept,
the notion of 
{\em context transitions} that we have introduced in the development is novel. 
It models the interactions between programs and contexts and may have potential use in 
game semantics~\cite{AM98,HO00}. 
We also notice that such context transitions (along with program transitions) conforms to the idea of 
rely-guarantee reasoning, which has been successfully applied in the verification of
concurrent programs~\cite{XRH97,VP07,Fen09}, and may suggest an alternative approach.

Although the entire development is based on an operational treatment, the technique is 
general enough to be adapted in other languages with linear type systems.  
Indeed, we show that our approach can be applied in a non-deterministic extension of 
the linear PCF based on Moggi's framework with monadic types, where trace equivalence also 
serves as a sound and complete characterization of linear contextual equivalence. 
The result particularly helps us to prove the equivalence of the two functions in the previous
example, as we can show that they are trace equivalent. 

One can probably employ Howe's approach when proving linear contextual equivalence 
in a deterministic language. 
While Howe's approach  applies to a wider variety of occasions, it is more involved;
our approach is much 
simpler because we take advantage of linearity in resource usage. 
Furthermore, in non-deterministic languages, simulation based
techniques fail to 
characterize linear contextual equivalence.

\ifLONG
\subsection{Outline}
The rest of the paper is organized as follows: Section~\ref{sec:linpcf} defines briefly 
a linear version of call-by-name PCF with a dual type system, as well as its operational semantics. 
In particular, a labeled transition system for the language is presented and 
the notion of trace equivalence is defined. 
In Section~\ref{sec:lin.cont} we introduce the notion of linear contextual equivalence and show that 
trace equivalence in linear PCF coincides with linear contextual equivalence. 
Section~\ref{sec:nlpcf} extends our approach in a non-deterministic circumstance with monadic types, 
where technical development follows the previous two sections, and we establish the coincidence 
between trace equivalence and linear contextual equivalence. 
With this result, we show that the two functions in the previous example are indeed equivalent
in linear contexts. 
Section~\ref{sec:conclu} concludes the paper.
\else
\vspace*{-1.2em}
\subsubsection{Outline.}
Section~\ref{sec:linpcf} defines a linear version of call-by-name PCF, as well as a labeled transition system.
The notion of trace equivalence is also defined. 
In Section~\ref{sec:lin.cont} we introduce the notion of linear contextual equivalence and show that 
it coincides with trace equivalence in linear PCF.
Section~\ref{sec:nlpcf} extends our approach in a non-deterministic circumstance with monadic types,
and Section~\ref{sec:conclu} concludes the paper.
\fi
\ifLONG
\else
Due to the page limit, most proofs 
 are omitted --- they can be found in the full version 
of the paper~\cite{DZ11-full}.
\fi

%% file: lpcf.tex
\section{The call-by-name linear PCF}
\label{sec:linpcf}

We start with a linear version of PCF (LPCF for short) with a call-by-name evaluation strategy. 
Types are given by the following grammar:
\[
\tau, \tau', \ldots ::= \tnat \mid \tbool \mid \tau \product \tau' \mid \tau \otimes \tau' 
\mid \tau \linto \tau' \mid \tau \to \tau' 
\]
Here $\tau \product \tau'$ and $\tau \tensor \tau'$ are usual product
and tensor product 
respectively. 
Linear functions will be given types in the form $\tau \linto
\tau'$. Following 
\cite{Plo93}, 
we choose to make intuitionistic function types $\tau \to \tau'$ primitive rather than introducing 
exponential types. 
The choice makes our technical development simpler but does not affect the heart of the approach
--- one can certainly express non-linear function types in terms of $\bang$-types, using Girard's 
decompositon: $\tau \to \tau' = \,!\tau \linto \tau'$, and adapt our
technique 
accordingly.

Terms are built up from constants (boolean and integer values plus integer operations and 
fix-point recursion) and variables, using the following constructs.
\[
\begin{array}{lrl@{\qquad}l}
e, e', \ldots 
& ::= & x & \mbox{Variables}
\\ 
& \mid & \mathtt 0 \mid \mathtt 1 \mid \mathtt 2 \mid \ldots & \mbox{Integers}
\\
& \mid & \csucc \mid \cpred \mid \czero & \mbox{Integer operations}
\\
& \mid & \lambda x \ldot e \mid e \, e' & \mbox{Abstractions and applications}
\\
& \mid & \ctrue \mid \cfalse & \mbox{Booleans}
\\
& \mid & \cif{e_1}{e_2}{e_3} & \mbox{Conditionals}
\\ 
& \mid & \cpair{e_1, e_2} \mid \cproj_i(e) & \mbox{Products and projections}
\\
& \mid & \cfix_\tau& \mbox{Fix-point recursions}
\\
& \mid & e_1 \otimes e_2 \mid \clet{x \otimes y}{e}{e'} & \mbox{Tensor products and projections}
\end{array}
\]

\ifLONG
Most of the language constructs are standard: 
the $\lambda$-abstraction $\lambda x.e$ defines a function, whose linearity will be judged by the type system, 
and the application $e\; e'$ applies the function $e$ to the argument $e'$; 
the conditional $\cif{e_1}{e_2}{e_3}$ evaluates like $e_2$ or $e_3$, according to whether the boolean term $e_1$ 
evaluates to $\ctrue$ or $\cfalse$;
$\cpair{e_1,e_2}$, $\cproj_1{e}$ and $\cproj_2{e}$ are normal products and corresponding projections;
the term $\cfix_{\tau}{e}$ represents the least fix-point of the function $e$. 
The tensor product and tensor projection are related to linearity --- the constructs actually force 
that no single component of a product can be discarded while the other is preserved. 
Tensor products are also useful for currying linear functions.  
\else
Most of the language constructs are standard. The tensor product and tensor projection are related to linearity.
\fi
\ifLONG

\else
\fi
Variables appearing in the $\lambda$-binder and the $\mathtt{let}$-binder (in tensor projections) 
are bound variables of LPCF programs. 
\ifLONG
We write $\FV(e), \FLV(e), \FNV(e)$ for the sets of, respectively, free variables, 
free linear variables, and free non-linear variables in term $e$.
We will not distinguish \emph{$\alpha$-equivalent} terms, which are terms syntactically identical 
up to renaming of bound variables. 
If $e$ and $e'$ are terms and $x$ is a variable, then $e[e'/x]$ denotes the term resulting from 
substituting $e'$ for all free occurrences of $x$ in $e$. 
More generally, given a list $e_1,...,e_n$ of terms and a list $x_1,...,x_n$ of distinct variables, 
we write $e[e_1/x_1,...,e_n/x_n]$ for the result of simultaneously substituting each term $e_i$ 
for free occurrences in $e$ of the corresponding variable $x_i$.
\else
\fi

A \emph{typing assertion} takes the form $\Gamma;\Delta\vdash e:\tau$, where $\Gamma$ and $\Delta$ 
are finite partial functions from variables to types, $e$ is a term, and $\tau$ is a type. 
We adopt the notation from dual intuitionistic linear logic~\cite{Bar96} by using $\Gamma$ 
and $\Delta$ to represent typing environments for, respectively, non-linear variables and 
linear variables. 
It is assumed that the codomains of the non-linear and linear typing environments are disjoint.
The \emph{type assignment relation} for the linear PCF consists of all typing assertions that can 
be derived from the axioms and rules in Figure~\ref{fig:type}, which are very standard. 
\ifLONG
The notation $\Gamma, x:\tau$ denotes the partial function which properly extends $\Gamma$ 
by mapping $x$ to $\tau$, so it is implicitly assumed that $x$ is not in the domain of $\Gamma$.
\else
\fi
We write $\prog(\tau) = \set{e \mid \emptyset ; \emptyset \vdash e : \tau }$ 
for the set of all closed programs of type $\tau$. 

\begin{figure}
\fbox{\parbox{\textwidth}{
\[
\begin{array}{c}
\begin{prooftree}
x : \tau \in \Gamma
\justifies 
\Gamma ; \emptyset\vdash x : \tau 
\end{prooftree}
\qquad
\begin{prooftree}
x : \tau \not\in \Gamma
\justifies 
\Gamma ; x : \tau \vdash x : \tau 
\end{prooftree}
\qquad
\prooftree
\justifies
\Gamma ; \emptyset \vdash \cfix_\tau : (\tau \to \tau) \to \tau
\endprooftree
\qquad 
\begin{prooftree}
i \in \set{\mathtt 0, \mathtt 1, \mathtt 2, \ldots }
\justifies 
\Gamma ; \emptyset \vdash i : \tnat
\end{prooftree}
\\[1.5em]
\begin{prooftree}
\justifies 
\Gamma ; \emptyset \vdash \csucc : \tnat \linto \tnat
\end{prooftree}
\qquad
\begin{prooftree}
\justifies 
\Gamma ; \emptyset \vdash \cpred : \tnat \linto \tnat
\end{prooftree}
\qquad
\begin{prooftree}
\justifies 
\Gamma ; \emptyset \vdash \czero : \tnat \linto \tbool
\end{prooftree}
\\[1.5em]
\begin{prooftree}
b \in \set{\ctrue, \cfalse}
\justifies 
\Gamma ; \emptyset \vdash b : \tbool
\end{prooftree}
\qquad
\begin{prooftree}
\Gamma; \Delta \vdash e_1 : \tbool
\quad 
\Gamma; \Delta' \vdash e_2 : \tau
\quad 
\Gamma; \Delta' \vdash e_3 : \tau
\justifies 
\Gamma ; \Delta, \Delta' \vdash \cif{e_1}{e_2}{e_3} : \tau
\end{prooftree}
\\[1.5em]
\begin{prooftree}
\Gamma ; \Delta \vdash e_i : \tau_i \; (i = 1, 2)
\justifies
\Gamma ; \Delta \vdash \cpair{e_1, e_2} : \tau_1 \product \tau_2
\end{prooftree}
\qquad
\begin{prooftree}
\Gamma ; \Delta \vdash e : \tau_1 \product \tau_2
\justifies
\Gamma ; \Delta \vdash \cproj_i (e) : \tau_i \; (i = 1, 2)
\end{prooftree}
\\[1.5em]
\begin{prooftree}
\Gamma ; \Delta_i \vdash e_i : \tau_1 \; (i = 1, 2)
\justifies
\Gamma ; \Delta_1, \Delta_2 \vdash e_1 \tensor e_2 : \tau_1 \tensor \tau_2
\end{prooftree}
\qquad 
\begin{prooftree}
\Gamma ; \Delta, x : \tau_1, y : \tau_2 \vdash e : \tau 
\quad
\Gamma ; \Delta' \vdash e' : \tau_1 \tensor \tau_2 
\justifies
\Gamma ; \Delta, \Delta' \vdash \clet{x \tensor y}{e'}{e} : \tau 
\end{prooftree}
\\[1.5em]
\begin{prooftree}
\Gamma, x : \tau ; \Delta  \vdash e : \tau'
\justifies 
\Gamma ; \Delta \vdash \lambda x \ldot e : \tau \to \tau'
\end{prooftree}
\qquad 
\begin{prooftree}
\Gamma ; \Delta  \vdash e : \tau' \to \tau
\quad 
\Gamma ; \emptyset \vdash e' : \tau' 
\justifies 
\Gamma ; \Delta \vdash e \, e'  : \tau'
\end{prooftree}
\\[1.5em]
\begin{prooftree}
\Gamma ; \Delta , x : \tau \vdash e : \tau'
\justifies 
\Gamma ; \Delta \vdash \lambda x \ldot e : \tau \linto \tau'
\end{prooftree}
\qquad 
\begin{prooftree}
\Gamma ; \Delta  \vdash e : \tau' \linto \tau
\quad 
\Gamma ; \Delta'\vdash e' : \tau' 
\justifies 
\Gamma ; \Delta, \Delta' \vdash e \, e'  : \tau'
\end{prooftree}
\end{array}
\]
\caption{LPCF typing rules}\label{fig:type}
}}
\end{figure}

\subsection{The operational semantics}\label{sec:red.sem}

We first define the notion of \emph{values} of LPCF. 
\[
\begin{array}{rl}
v, v', \ldots ::= & \csucc \mid \cpred \mid \czero \mid \ctrue \mid \cfalse 
\mid \mathtt 0 \mid \mathtt 1 \mid \mathtt 2 \mid \ldots 
\\
\mid & \cfix_\tau \mid \cpair{e, e'} \mid e \tensor e' \mid \lambda x \ldot e 
\end{array}
\]
These are also {\em canonical forms} of LPCF 
terms.

The one-step reduction 
$\reduce$ between terms is inductively
defined by the 
axioms
\[
\begin{array}{l}
(\lambda x . e) e' \; \reduce \; e [ e' / x ]
\\
\cfix_\tau \, e  \; \reduce \; e (\cfix_\tau \, e)
\\
\csucc \, n \; \reduce \; n+1,  \; 
\mbox{where } n \in \set{\mathtt 0, \mathtt 1, \mathtt 2, \ldots}
\\
\cpred \, 0 \; \reduce \; 0
\\
\cpred \, n \; \reduce \; n-1,  \; 
\mbox{where } n \in \set{\mathtt 1, \mathtt 2, \ldots}
\\
\czero \, \mathtt 0 \; \reduce \; \ctrue 
\\
\czero \, n \; \reduce \; \cfalse, \; \mbox{where } n \in \set{\mathtt 1, \mathtt 2, \ldots}
\\
\cif{\ctrue}{e_1}{e_2} \; \reduce \; e_1 
\\
\cif{\cfalse}{e_1}{e_2} \; \reduce \; e_2
\\
\cproj_i \cpair{e_1, e_2} \; \reduce \; e_i, \; (i = 1, 2)
\\  
\clet{x \tensor y}{e_1 \tensor e_2}{e} \; \reduce \; e [e_1 / x, e_2 / y]
\end{array}
\]
together with the structural rule
\[
\prooftree
e_1 \reduce e_2
\justifies 
\evcontext[e_1] \reduce \evcontext[e_2]
\endprooftree
\]
where $\evcontext$ is the evaluation context generated by the grammar
\[
\begin{array}{rl}
\evcontext ::= & \hole \mid \csucc(\evcontext) \mid \cpred(\evcontext) \mid \czero(\evcontext) 
\mid \evcontext \, e \mid \cif{\evcontext}{e_1}{e_2} \\
\mid & \cproj_i(\evcontext) \mid \clet{x \tensor y}{\evcontext}{e} 
\end{array}
\]
We often call a term $\evcontext[x]$ an evaluation context, 
if $x$ is the {\em only}
free variable of the term.

The operational semantics that we define for LPCF is essentially a {\em call-by-name} evaluation. 
Although our later development depends on the operational semantics, it does not really matter 
whether the evaluation strategy is call-by-name or call-by-value --- one can easily 
adapt our approach to a call-by-value semantics. 
The only crucial point is that we should not allow the following forms
of evaluation contexts:
$$\cpair{\evcontext, e}, \ \cpair{e, \evcontext}, \ 
\cif{e}{\evcontext}{e'}, \ \cif{e}{e'}{\evcontext}.$$ 
This is because these contexts adopt syntactically duplicated linear variables without breaking 
linearity restriction, 
hence if we substitute a reducible term for such
a variable, which makes multiple 
copies of the term in the context, 
then one of them may be reduced while 
all other copies remain unchanged. We shall see how this fact affects our approach in more detail.
Indeed, such restriction over evaluation contexts conforms to the semantics of linearity --- 
as long as a program is allowed to be ``used'' only once, it should not be reduced for multiple 
times, hence we can safely adopt such evaluation restriction in languages with linear types.

It is clear that LPCF terms in canonical form do not reduce. The following proposition also 
shows that every closed non-reducible term must be in the canonical form. 
We write $e \not\reduce$ when there does not exist a term $e'$ such that $e \reduce e'$, 
and $\reduce^*$ denotes the reflexive transitive closure of $\reduce$.

\begin{proposition}\label{prop:canonical}
If $e$ is a closed term and $e \not\reduce$, then $e$ must be in the canonical form.
\end{proposition}
\ifLONG
\begin{proof}
We prove by induction on the structure of $e$. Below is the analysis for non-canonical forms:
\begin{itemize}
\item $e \equiv \cif{e'}{e_1}{e_2}$. Here $e'$ must be closed and not reducible (otherwise the whole 
  term can be reduced since $\cif{\hole}{e_1}{e_2}$ is an evaluation context). By induction 
  $e'$ must be canonical, i.e., either $\ctrue$ or $\cfalse$, but in both cases, the original 
  term can be reduced.
\item $e \equiv \cproj_i(e')$. Here $e'$ must be closed and not reducible (since $\cproj_i \hole$ is 
  an evaluation context), and by induction, must be the canonical form $\cpair{e_1, e_2}$, 
  which makes the original term reducible.
\item $e \equiv \clet{x \tensor y}{e'}{e''}$. Here $e'$ must be closed and not reducible, 
  and by induction, must be the canonical form $e_1 \tensor e_2$, which makes the original 
  term reducible.
\item $e \equiv e' \, e''$. Here $e'$ must be closed and not reducible, and by induction, must be 
  canonical: if $e'$ is an abstraction or a fix-point, then the whole term can be reduced; 
  if $e' \in \set{\csucc, \cpred, \czero}$, then $e''$ must be canonical, which will be an 
  integer, hence the whole term can be reduced too.
\qed
\end{itemize} 
\end{proof}
\else
\fi

Evaluation in LPCF is deterministic and preserves typing.
\ifLONG
\begin{lemma}\label{lem:linvar}
For every well-typed term $e$, if $e \reduce e'$, then $\FLV(e') = \FLV(e)$.
\end{lemma}
\begin{proof}
By rule induction on the derivation of $e \reduce e'$.
\qed
\end{proof}
\else
\fi

\begin{proposition}[Subject reduction]\label{prop:subj}
If $\Gamma; \Delta \vdash e : \tau$ and $e\reduce e'$, 
then $\Gamma; \Delta \vdash e': \tau$.
\end{proposition}
\ifLONG
\begin{proof}
A routine exercise.
\qed
\end{proof}
\else 
\fi

\begin{proposition}[Determinacy]\label{prop:det}
\begin{enumerate}
\item If $e\reduce^* v\not\reduce$ and $e\reduce^* v'\not\reduce$ then $v=v'$.
\item 
Every well-typed term either converges or all of its reduction do not terminate.
\end{enumerate}
\end{proposition}
\ifLONG
\begin{proof}
It suffices to prove that the evaluation is deterministic, that is, there is at most one
reduction rule that applies in any situation. This can be proved by
induction on the structure of terms.
\qed
\end{proof}
\else
\fi

Because the reduction is deterministic in LPCF, for any closed term $e$, we say $e$ 
\emph{converges} and write $e \Reduceto$ if it reduces to a value.
Conversely, we say $e$ \emph{diverges} if the reduction of $e$ does not terminate 
and we write $e\Uparrow$.
We also define a specific class of terms $\infini_\tau \defeq \cfix_\tau (\lambda x \ldot x)$, 
to represent non-terminating programs.

\subsection{A labeled transition system for LPCF}\label{sec:lts}

In~\cite{Gor95}, Gordon defines explicitly a labeled transition system in order to illustrate 
the applicative bisimulation technique in PCF. 
We follow this idea to define a labeled transition system for LPCF, upon which we can define 
the notions of traces and trace equivalence and develop our framework.

Transition rules are listed in Figure~\ref{fig:trans}: 
we make the typing of terms explicit in the rules as the type system plays an important role in LPCF. 
\begin{figure}
\fbox{\parbox{\textwidth}{
\[
\begin{array}{c}
\prooftree
c \in \set{\ctrue, \cfalse, \mathtt 0, \mathtt 1, \mathtt 2, \ldots}
\justifies
c \transto{c} \infini
\endprooftree
\\[2em]
\prooftree
\Gamma; \Delta \vdash \lambda x \ldot e : \tau
\quad 
\emptyset; \emptyset \vdash e' : \tau'
\quad 
\tau \equiv \tau' \linto \tau'' \mbox{ or } \tau' \to \tau''
\justifies
\lambda x \ldot e \transto{@e'} e [e'  / x]
\endprooftree
\\[2em]
\prooftree
\Gamma; \Delta \vdash \cpair{e_1, e_2} : \tau_1 \product \tau_2
\justifies
\cpair{e_1, e_2} \transto{\cproj_i} e_i
\endprooftree
\\[2em]
\prooftree
\Gamma; \Delta \vdash e_1 \tensor e_2 : \tau_1 \tensor \tau_2
\qquad 
\emptyset; x: \tau_1, y:\tau_2 \vdash e : \tau
\justifies
e_1 \tensor e_2 \transto{\tensor e} e [e_1 / x, e_2 / y]
\endprooftree
\\[2em]
\begin{prooftree}
e \reduce e'' \qquad e'' \xrightarrow{\,\alpha\,} e'
\justifies 
e \xrightarrow{\,\alpha\,} e'
\end{prooftree}
\end{array}
\]
\caption{Labeled transition system for LPCF}
\label{fig:trans}
}}
\end{figure}

The last rule in Figure~\ref{fig:trans} says that term reductions are considered as 
internal transitions --- external transitions are labeled by {\em actions}. 
Note that in the sequel, we shall write $e \transto{\alpha} e'$ for a single external 
transition without preceding internal transitions, and make internal transitions explicit 
when $e \reduce \cdots \reduce \transto{\alpha} e'$.

Intuitively, external transitions represent the way terms interact with environments (or contexts). 
For instance, a $\lambda$-abstraction can ``consume'' (application of itself to) a term, 
which is supplied by the environment as an argument, and forms a $\beta$-reduction.
The first rule says that, what an integer or boolean constant can provide to the environment 
is the value of itself, and after that it can no more provide any information, 
hence no external transitions can occur any more.
We represent this by a transition, labeled by the value of the constant, 
into a non-terminating program $\pmb{\Omega}$ of appropriate type. 

It should be noticed that transitions are defined in general for LPCF terms, including open terms, 
but they never introduce new free variables. This is particularly true for 
$@$- and $\tensor$-transitions according to their typing premises.

Let $s$ be a finite sequence of actions 
$\alpha_1 \alpha_2 \ldots \alpha_n$ ($n \geq 1$). 
We write $e \transto{s}$ if there exist terms $e_1, e_2, \ldots , e_n$ such that 
$e \reduce^* \transto{\alpha_1} e_1 \reduce^* \transto{\alpha_2} e_2 \dots \reduce^* \transto{\alpha_n} e_n$
(the entire sequence including term reductions is called the {\em full sequence} of $s$). 
An action sequence $s$ is {\em a trace of $e$} if $e \transto{s}$, and we write 
$\trace(e)$ for the set of all traces of $e$, i.e., $\trace(e) \defeq \set{s \mid e \transto{s}}$.
We also write $\alpha \cdot s$ and $s_1 \cdot s_2$ for the traces obtained by, respectively, 
prefixing trace $s$ with an action $\alpha$ and concatenating $s_1$ and $s_2$.

Given two traces $s_1$ and $s_2$, we say $s_1$ is a \emph{subtrace} of $s_2$ if $s_1$ is a prefix of $s_2$ when they are viewed as strings.  
A trace of a LPCF term $e$ is \emph{maximal} if it is not a subtrace of any other trace of $\trace(e)$.
A \emph{computational} trace is a maximal trace of the form $s \cdot c$, where $c$ is a boolean or integer constant.
\ifLONG
In other words, a computational trace ends with some observable value, while a non-computational trace may end with an action in the forms $@e,\ \cproj_i,\ \otimes e$ or $\tcomp$.
\else
\fi

\ifLONG

\else
\fi
The empty trace, denoted by $\nulltrace$, can be taken by any program. 
Meanwhile, if $\nulltrace$ is the only trace that a term can take, which means 
the term cannot take any external action, then the term must diverge, i.e, 
$\trace(e) = \set{\nulltrace} \mbox{ iff } e \diverge$.

We define the {\em trace preorder} $\preord^T$ 
between terms: $e_1 \preord^T e_2$ iff 
$\trace(e_1) \subseteq \trace(e_2)$. 
Two terms $e_1$ and $e_2$ are trace equivalent, written $e_1 \simeq^T e_2$, iff $e_1 \preord^T e_2$ and 
$e_2 \preord^T e_1$.

\ifLONG
\begin{lemma}\label{lem:tr.type}
\begin{enumerate}
\item 
If $\trace(e_1) = \trace(e_2) \not= \set{\nulltrace}$,
  then $e_1, e_2$ must have the same type.
\item Let $e_1,\, e_2$ be two terms of the same type. For any trace $s$, if $e_1\transto{s} e'_1$ and $e_2\transto{s} e'_2$, then $e'_1$ and $e'_2$ also have the same type.
\item If $e \reduce e'$, then $e' \preord^T e$.
\end{enumerate}
\end{lemma}
\begin{proof}
The first statement can be proved by contradiction; the second one is proved by induction on the length of $s$; the third one is a direct consequence of the definition of trace preorder.
\qed
\end{proof}
\else
\fi

\section{Linear contextual equivalence}\label{sec:lin.cont}

Defining a context in a language with linear types must be treated carefully, since holes can hide 
bound variables and consequently breaks the typing if the variable is linear~\cite{Bie00}.
We choose to replace the context hole by an explicit free variable and restrict attention to 
equivalence between closed terms, so as to avoid extra syntactic machinery.

Intuitively, a linear context is a context where programs under 
examination will be evaluated and 
used {\em exactly once}\footnote{It is more general to consider
  \emph{affine} contexts where programs are 
executed {\em at most} once, but in the current paper we refrain from going that far and
leave it as future work. 
}.
In a linear functional language, we can formalize it by a restricted notion of contexts:
a {\em linear context} $\context_{x: \tau}$ in LPCF is a 
term with a single linear variable $x$
and no non-linear variables, i.e., $\emptyset; x : \tau \vdash \context_{x: \tau} : \sigma$.
We often omit the variable and type subscription when it is clear from the texts or irrelevant.

\begin{definition}[Linear contextual equivalence]\label{def:preordC}
We write $e_1 \preord^C e_2$ for $e_1, e_2 \in \prog(\tau)$, if 
$\context[e_1 / x] \Reduceto$ implies $\context[e_2 / x] \Reduceto$ 
for all linear context $\context_{x:\tau}$.
The relation $\preord^C$ is called the {\em linear contextual preorder} between closed programs. 
{\em Linear contextual equivalence $\simeq^C$} is defined as the symmetrization of $\preord^C$:
$e_1 \simeq^C_\tau e_2$ iff $e_1 \preord^C_\tau e_2$ and $e_2 \preord^C_\tau e_1$.
\end{definition}

In~\cite{BPR00}, the definition of {\em ground contextual equivalence} (Definition~2.1) 
says that contexts must be of exponential types, because they are necessary 
for a program to adopt recursions in their type system. In LPCF non-linear function types 
are primitive, with no exponential types, and the type for fix-point operator indicates that 
recursions must be taken within non-linear functions. Hence, the above definition 
admits the requirement of
 the definition of ground contextual equivalence in~\cite{BPR00}.

\ifLONG
\begin{lemma}
Let $\context_1, \context_2$ be two linear contexts such that 
$\emptyset; x: \tau \vdash \context_1 : \sigma$ and $ \emptyset; y : \sigma \vdash \context_2 : \sigma'$, 
then $\context_2[\context_1 / y]$ is also a linear context.
\end{lemma}
\begin{proof}
It can be shown that $\emptyset; x: \tau \vdash \context_2[\context_1 / y] : \sigma'$.
\qed
\end{proof}
\else
\fi

\subsection{Linear context transitions}

Corresponding to the transition system for terms, we also define transitions for linear contexts,
which only occur in evaluation contexts: 
\[
\begin{array}{rcl}
\context[\cif{x}{e_1}{e_2}/y] & \lintransto{\ctrue} & \context[e_1 / y]
\\
\context[\cif{x}{e_1}{e_2}/y] & \lintransto{\cfalse} & \context[e_2 / y]
\\
\context[\cpred(x)/y] & \lintransto{n} & \context[n'/y] \; (n = n' + 1  \mbox{ or } n = n' = 0)
\\
\context[\csucc(x)/y] & \lintransto{n} & \context[n'/y] \; (n' = n + 1) 
\\
\context[\czero(x)/y] & \lintransto{n} & \context[\ctrue/y] \; (\mbox{if } n = 0) 
\\
\context[\czero(x)/y] & \lintransto{n} & \context[\cfalse/y] \; (\mbox{if } n \neq 0) 
\\
\context[\cproj_i(x) / y] & \lintransto{\cproj_i} & \context_y \; (i = 1, 2)
\\
\context[x \, e / y] & \lintransto{@e} & \context_y 
\\
\context[\clet{z_1 \tensor z_2}{x}{e}/y] & \lintransto{\tensor e} & \context_y
\end{array}
\]
Linear context transitions represent the way a context interact with programs under testing.
A linear context transition often eliminates the free variable in the context or transforms it 
into another variable of a different type (in which case we often use a variable with a different 
name for the sake of clarity), which indicates that a reduction can occur involving 
both the candidate program and (a subterm of) the context.  

\ifLONG
Linear context transitions do not necessarily transform a linear context into another linear context 
--- linear contexts can also be transformed into closed terms, which do not contain any free variables.
This particularly happens when the program under testing is a boolean or integer constant, which, 
after transition, cannot provide any information to the context. 
\else 
\fi

Notice that linear contexts themselves are LPCF terms, so they can also take normal transitions as 
defined in Figure~\ref{fig:trans}. We have used explicitly distinguished notations for the two kinds 
of transitions.
\begin{lemma}[Transition lemma]
\label{lem:lc-trans}
Fro every linear context $\context_{x:\tau}$ and LPCF program $e \in \prog(\tau)$
such that $\context[e/x] \not\reduce$, 
a transition from $\context[e/x]$ must be either of the two forms:
\begin{itemize}
\item $\context[e/x] \transto{\alpha} \context'[e/x]$ with $\context \transto{\alpha} \context'$;
\item $\context \equiv x$ and $\context[e/x] \equiv e \transto{\alpha} e'$.
\end{itemize}
\end{lemma}
\ifLONG
\begin{proof}
Since $\context[e/x] \not\reduce$, it must be in the canonical form, then $\context$ must be 
one of the forms: $x$, $\context_1 \tensor e'$, $e' \tensor \context_1$, 
$\cpair{\context_1, \context_2}$, $\lambda y \ldot \context_1$,
where $e'$ is a closed term and $\context_1, \context_2$ are linear contexts with free variable $x$.

It is clear that if $\context \equiv x$, the transition must be of the second form. 
In all other forms, it can be easily checked that the transition will be of the first form, with the context 
$\context$ itself being transformed into another term with the free linear variable $x$, which forms another 
linear context.
\qed
\end{proof}
\fi

\subsection{Linear context reductions}

%
Reductions of linear contexts filled with programs can be classified into several forms, called 
{\em linear context reductions} (LCR for short), which characterize the interaction between 
linear contexts and programs.
\begin{definition}[Linear context reduction]
Let $\context_{x: \tau}$ be a linear context and $e \in \prog(\tau)$ be a LPCF program.
A reduction of $\context[e / x]$ (if it is reducible) is called a {\em linear context reduction} 
if it is either of the following forms: 
\begin{itemize}
\item $\context [e / x] \reduce \context' [e / x]$, if $\context \reduce \context'$;
\item $\context [e / x] \reduce \context [e' / x]$, if $\context$ is an evaluation context, 
  and $e \reduce e'$;
\item $\context [e / x] \reduce \context' [e' / y]$, if $\context$ is an evaluation context,
  $e \not \reduce$, and $\context \lintransto{\alpha} \context'$, $e \transto{\alpha} e'$ 
  for some external action $\alpha$.
\end{itemize}
\end{definition}
We often write $\context[e/x] \intransto{\alpha} \context'[e'/y]$ for the third form of 
linear context reduction, indicating explicitly that the transitions involved are labeled by $\alpha$.

\ifLONG
Linear context reductions are closed under linear evaluation contexts:
\begin{lemma}\label{lem:eval}
Let $\context_1, \context_2$ be two linear contexts such that 
$\emptyset; x: \tau \vdash \context_1 : \sigma$ and $ \emptyset; y : \sigma \vdash \context_2 : \sigma'$,
and $\context_2$ also an evaluation context.
\begin{enumerate}
\item If $\context_1 \transto{\alpha} e$, then $\context_2[\context_1/y] \transto{\alpha} \context_2[e/y]$.

\item If $\context_1[e/x] \reduce e'$ is a linear context reduction, 
then $\context_2[\context_1[e/x]/y] \reduce \context_2[e'/y]$ is also a linear context reduction. 
\end{enumerate}
\end{lemma}
\begin{proof}
Direct consequence of the definition of linear context transitions.
\qed
\end{proof}
\else
\fi

The so-called {\em linear context reduction lemma} below says that, 
the reduction of a linear context filled with a program in LPCF must be a linear context reduction.
\ifLONG
This is the core lemma of proving precongruence of trace equivalence w.r.t. linear contexts. 
\else 
\fi

\begin{lemma}[Linear context reduction lemma]
\label{lem:lcr}
For every linear context $\context_{x:\tau}$ and LPCF program $e \in \prog(\tau)$, 
if $\context[e / x]$ is reducible, then $\context[e/x] \reduce$ must be a linear context reduction.
\end{lemma}
\ifLONG
\begin{proof}
We prove by induction on the structure of the linear context. 

\begin{itemize}
\item $\context$ cannot be any constant since it must contain a linear free variable. And it cannot be 
  a normal product, a tensor product or an abstraction, as all these forms 
  cannot be reduced any more, no matter what $e$ is.
\item The simplest linear context $x$ is an evaluation context. If $e$ can be reduced, 
  then it is the second case.
\item $\context \equiv \cif{\context'}{e_1}{e_2}$, where $\context'$ is another linear context.
  If $\context'[e / x]$ can be reduced, by induction, it must be either of the following cases:
  \begin{itemize}
  \item $\context'[e / x] \reduce \context''[e / x]$ and $\context' \reduce \context''$,
    then we have that 
    $$\context[e/x] \reduce \cif{\context''[e/x]}{e_1}{e_2}$$
    with 
    $\context \reduce \cif{\context''}{e_1}{e_2}$.
  \item $\context'[e/x] \reduce \context'[e'/x]$, $e \reduce e'$, and $\context'$ is an evaluation context, 
    then $\context$ is also an evaluation context, hence $\context[e/x] \reduce \context[e'/x]$.
  \item $\context'[e/x] \reduce \context''[e'/x]$ and $\context' \lintransto{\alpha} \context''$, 
    $e \transto{\alpha} e'$ for some action $\alpha$, then 
    $$\context \lintransto{\alpha} \cif{C''}{e_1}{e_2}$$
    and $\context[e / x]$ can take a similar reduction.
  \end{itemize}
  If $\context'[e / x]$ cannot reduce, then it is a canonical boolean term, which is either $\ctrue$ 
  or $\cfalse$, 
  and the only possibility of $\context'$ is the simplest case $x$, with $e$ being a boolean constant. 
  In this case both $\context$ and $e$ can take the transition $\transto{\ctrue}$ or $\transto{\cfalse}$, 
  and the reduction of $\context[e / x]$ falls into the third case.
\item $\context \equiv \cif{e'}{\context_1}{\context_2}$, where by typing, both $\context_1$ and $\context_2$ 
  are linear contexts. If $e'$ can be reduced ($e' \reduce e''$), then $\context[e/x]$ will reduce to 
  $\cif{e''}{\context_1[e/x]}{\context_2[e/x]}$, which is still a linear context. If $e'$ cannot be reduced, 
  then it must be a boolean constant since it must be canonical, then $\context[e/x]$ will reduce to 
  either $\context_1[e/x]$ or $\context_2[e/x]$.  Both reductions are the first form of {LCR}.
\item $\context \equiv \cproj_i (\context')$, where $\context'$ is a linear context. If $\context'[e/x]$ itself 
  can be reduced, then by induction, it must be in one of the three forms of linear context reduction. 
  In each case, it is easy to see that $\context[e/x]$ will take the same form of reduction. 
  
  If $\context'[e/x]$ is not reducible, then it must be of the form $\cpair{\_, \_}$. There are two cases 
  \begin{itemize}
  \item $\context' \equiv \cpair{\context_1', \context_2'}$, where both $\context_1'$ and $\context_2'$ 
    are linear contexts, then $\context[e/x] \reduce \context_i[e/x]$, which is the first form of 
    linear context reduction.
  \item $\context' \equiv x$ and $e \equiv \cpair{e_1, e_2}$. Now both $\context$ and $e$ can take 
    the transition $\transto{\cproj_i}$: $\context \equiv \cproj_i(x) \lintransto{\cproj_i} y$,
    $e \equiv \cpair{e_1, e_2} \transto{\cproj_i} e_i$ and $\context[e/x] \reduce e_i = y[e_i/y]$. 
    This is the third form of linear context reduction.
  \end{itemize}
\item $\context \equiv \clet{y_1 \tensor y_2}{\context'}{e'}$. 
  If $\context'$ is reducible, by induction, the reduction of $\context'[e/x]$ must be a linear 
  context reduction, then the reduction of $\context[e/x]$ will be a linear context reduction of 
  the same form.
  
  If $\context'[e/x]$ is not reducible, then there are two cases:
  \begin{itemize}
  \item $\context' \equiv \context'' \tensor e''$ or $\context' \equiv e'' \tensor \context''$,
    where $e''$ is a closed term and $\context''$ is a linear context. 
    Consider the first case without losing generality. $\context[e/x]$ will reduce to 
    $e'[\context''[e/x] / y_1, e'' / y_2]$. It is easy to check that $e'[e''/y_2]$ is also a linear context, 
    then so is $e'[e''/y_2, \context'' / y_1]$, so the reduction is a linear context reduction of 
    the first form.
  \item $\context' \equiv x$ and $e \equiv e_1 \tensor e_2$. Now both $\context$ and $e$ can take a 
    $\transto{\tensor e'}$ transition: 
    $\context \equiv \clet{y_1 \tensor y_2}{x}{e'} \lintransto{\tensor e'} z$, 
    $e \equiv e_1 \tensor e_2 \transto{\tensor e'} e'[e_1/y_1, e_2/y_2]$, and 
    $\context[e/x] \reduce e'[e_1/y_1, e_2/y_2] = z [e'[e_1/y_1, e_2/y_2]/z]$.
  \end{itemize}
\item $\context \equiv \clet{y \tensor z}{e'}{\context'}$. It
  is clear that $e'$ is a closed term and $\FLV(\context') = \set{x, y, z}$.
  If $e' \reduce e''$ reduces, then $\context[e/x] \reduce \clet{y \tensor z}{e''}{\context'[e/x]}$. 
  Otherwise, $e'$ must be $e_1' \tensor e_2'$, then 
  $\context[e/x] \reduce \context'[e/x, e_1'/y, e_2'/z]$, with $\context$ reducing to 
  $\context'[e_1'/y, e_2'/z]$, which is a linear context.
\item $\context \equiv \context' \, e'$.
  Because $\context'$ is a linear context, by induction, if $\context'[e/x]$ can be reduced, then 
  it must be a linear context reduction. As $\context'\,e'$ is an evaluation context, $\context[e/x]$ 
  will take the same form of linear context reduction as $\context'[e'/x]$.

  If $\context'[e/x]$ cannot be reduced, then it must be an abstraction. There are two cases:
  \begin{itemize}
  \item  $\context' \equiv \lambda y \ldot \context''$ and $x \in \FLV(\context'')$, then 
    $\context \equiv (\lambda y \ldot \context'')e' \reduce \context''[e'/y]$ and it is easy to check that 
    $\context''[e'/y]$ is a linear context since $e'$ is closed, 
    hence $\context[e/x] \equiv (\lambda y \ldot \context''[e/x]) e' \reduce \context''[e'/y][e/x]$, 
    which is the first form of linear context reduction. 
  \item $\context' \equiv x$ and $e \equiv \lambda y \ldot e''$ is an abstraction, then 
    $\context \equiv x \, e' \lintransto{@e'} z$ (with $z$ being a fresh linear variable, hence a linear context),
    $e \equiv \lambda y \ldot e'' \transto{@e'} e''[e'/y]$, and 
    $\context[e'/x] \equiv (\lambda y \ldot e'') e' \reduce e''[e'/y] \equiv z[e''[e'/y]/z]$.
  \end{itemize}
\item $\context \equiv e' \, \context'$.
  If $e' \reduce e''$, then $\context[e/x] \reduce e''(\context'[e/x])$ with 
  $\context \reduce e'' \, \context'$.
  If $e'$ cannot be reduced, then it must be canonical, which is either an abstraction or a constant. 
  Because $\context'$ contains a linear variable, according to the typing system, the type of $e'$ 
  can only be a linear function type. 
  \begin{itemize}
  \item If $e' \equiv \lambda y \ldot e''$, 
    $\context[e/x] \equiv (\lambda y \ldot e'') (\context'[e/x]) \reduce e''[\context'[e/x] / y] 
    = e''[\context'/y][e/x]$. Also $\context \reduce e''[\context'/y]$. 
    Because $y$ is a free linear variable in $e''$, $e''$ is indeed a linear context, so is $e''[\context'/y]$.
  \item If $e'$ is a constant, because its type must be a linear function type, so it can only be one 
    of $\set{\cpred, \csucc, \czero}$. In any case, $e' \, \context'$ is an evaluation context. 
    If $\context'[e/x]$ reduces, then by induction it must be a linear context reduction, hence 
    $e' \, \context[e/x]$ can reduce and is a linear context reduction of the same form as of $\context'[e/x]$.
    If $\context'[e/x]$ cannot reduce, it must be canonical, i.e., an integer $n$, then $\context' \equiv x$ 
    and $e \equiv n$. Now both $\context$ and $e$ can take a $\transto{n}$ transition and 
    $\context[e/x]$ will reduce to another integer or a boolean constant, depending on which constant 
    $e'$ is.
    \qed
  \end{itemize}
\end{itemize}
\end{proof}
\fi

The linear context reduction lemma is not true if the context is not linear 
or the language does not have linear types at all, because duplicated use of programs 
in the context will adopt reductions that cannot be characterized by LCR, particularly 
when the program itself is reducible, i.e., $\context[e/x] \reduce \context[e'/x]$ is not 
true when $e \reduce e'$ and $\context$ makes multiple copies of $e$.
The reduction strategy also interferes, as we have mentioned when defining the operational semantics:
introducing improper evaluation contexts like $\cpair{\evcontext, e}$ breaks the linear context
lemma, for the same reason as using non-linear contexts. 

\subsection{Soundness and completeness of trace equivalence}

We show that in LPCF, the trace preorder relation is precongruent with respect to linear contexts. 
It then enables us to show that trace equivalence actually coincides with linear contextual equivalence. 

\ifLONG
The following theorem says that trace preorder in LPCF is a precongruence relation with respect to 
linear contexts. As LPCF is a deterministic language, the proof can be done by induction on 
(the length of) traces. 
\begin{theorem}[Linear precongruence of $\preord^T$]
\label{thm:precong-lpcf}
Trace preorder $\preord^T$ is a precongruence with respect to linear contexts, i.e., 
$e_1 \preord^T e_2$ implies that $\context[e_1 / x] \preord^T \context[e_2 / x]$ 
for all linear contexts $\context_x$.
\end{theorem}
\begin{proof}
According to the definition of $\preord^T$, it suffices to show that, for any action sequence $s$, 
if $\context [e_1 / x] \transto{s}$, then $\context [e_2 / x] \transto{s}$. 
We prove by induction on the length of $\context[e_1 /x ] \transto{s}$ (note that the 
transition includes internal transitions, i.e., term reductions). The base case is trivial.

We distinguish two cases.
\begin{itemize}
\item $\context[e_1 / x] \reduce e \transto{s}$. By the linear context lemma, the reduction must be 
  a linear context reduction, which is one of the following cases:
\begin{itemize}
\item $e \equiv \context'[e_1 / x]$ where $\context \reduce
  \context'$. It holds that 
  $\context[e_2 / x] \reduce \context'[e_2 / x]$. 
  By induction, $\context'[e_2/x] \transto{s}$ since $\context'[e_1/x] \transto{s}$, 
  hence $\context[e_2/x] \reduce \context'[e_2/x] \transto{s}$, i.e., $\context[e_2/x] \transto{s}$.
\item $e \equiv \context[e_1' / x]$ where $e_1 \reduce e_1'$.
  We immediately have $e_1' \preord^T e_1 \preord^T e_2$ and by induction, 
  $\context[e_2 / x] \transto{s}$ because $\context[e_1' / x] \transto{s}$.
\item $e \equiv \context_y'[e_1' / y]$ where $\context \lintransto{\alpha} \context_y'$ 
  and $e_1 \transto{\alpha} e_1'$. Since $e_1 \preord^T e_2$ and the
  transitions are deterministic, we have $e_2 \transto{\alpha} e_2'$ and 
  $e_1' \preord^T e_2'$. It is clear that $e_2 \transto{\alpha} e_2'$ must be of the form 
  $e_2 \reduce^* e_2'' \transto{\alpha} e_2'$, where $e_2'' \not\reduce$. By the definition of 
  linear context reduction, $\context$ must be an evaluation context, hence 
  $\context[e_2 / x] \reduce^* \context[e_2'' / x] \reduce \context_y'[e_2' / y]$, 
  and by induction, $\context_y'[e_1' / y] \transto{s}$ implies $\context_y'[e_2' / y] \transto{s}$. 
\end{itemize}
\item $\context[e_1 / x] \transto{\alpha} e \transto{s'}$. By Lemma~\ref{lem:lc-trans}, the first transition 
  has two forms:
\begin{itemize}
\item $\context \transto{\alpha} \context'$ and $e \equiv \context'[e_1/x]$. By induction, $\context'[e_1/x] \transto{s'}$ implies $\context'[e_2/x] \transto{s'}$. It follows that $\context[e_2/x]\transto{\alpha}\context'[e_2/x] \transto{s'}$. 
\item $\context \equiv x$ and $e_1 \transto{\alpha} e_1' \equiv e$. Then $e_1 \preord^T e_2$ implies that 
  $\context[e_2/x] \equiv e_2 \transto{\alpha} \transto{s'}$.
\qed
\end{itemize}
\end{itemize}
\end{proof}

However, the above proof does not apply in non-deterministic languages as trace preorder does not 
conform to induction in general. 
\else
In a language with deterministic reduction strategy like LPCF, normally we can prove the precongruence 
of trace preorder by induction on (length of) traces, however such a proof method does not apply 
in non-deterministic languages as trace preorder does not conform to induction in general. 
\fi
We supply in this section a more general proof for proving linear precongruence of trace preorder, 
by exploiting the intrinsic features of linear contexts. 

For every linear context $\context_{x: \tau}$ and LPCF program $e \in \prog(\tau)$, 
if $\context[e/x] \transto s$ and $e \transto t$, we define $t$ to be the {\em context trace} 
w.r.t. $\context$ and $s$ (also written as $(\context, s)$-trace), 
inductively on the full sequence of $s$, if:
\begin{itemize}
\item $t = \nulltrace$ when $s$ is empty;
\item $t$ is the context trace of $e'$ w.r.t. $\context$ and $s$ when $\context[e/x] \reduce \context[e'/x]$
  with $e \reduce e'$;
\item $t$ is the context trace of $e$ w.r.t. $\context'$ and $s$ when $\context[e/x] \reduce \context'[e/x]$
  with $\context \reduce \context'$;
\item $t = \alpha \cdot t'$ and $t'$ is the context trace of $e'$ w.r.t. $\context'$ and $s$ 
  when $\context[e/x] \intransto{\alpha} \context'[e'/x]$;
\item $t$ is the context trace of $e$ w.r.t. $\context'$ and $s'$ 
  when $\context[e/x] \transto{\alpha} \context'[e/x]$ with $\context \transto{\alpha} \context'$ 
  and $s = \alpha \cdot s'$;
\item $t = s$ when $\context \equiv x$.
\end{itemize}
%

\begin{lemma}\label{lem:ctx-trace}
For every linear context $\context_{x: \tau}$ and LPCF program $e \in \prog(\tau)$, 
if $\context[e/x] \transto s$, then $e$ has a context trace w.r.t. $\context$ and $s$. 
\end{lemma}
\ifLONG
\begin{proof}
The definition of context trace is solid by Lemma~\ref{lem:lcr} and Lemma~\ref{lem:lc-trans}, 
hence it is always feasible to construct the $(\context, s)$-trace from the full sequence of $s$
--- the definition indeed gives the construction.
\qed
\end{proof}
\fi
We also write $\context[e/x] \transto{(s, t)}$ when $t$ is a context trace of $e$ w.r.t. $\context$ and $s$.

\begin{lemma}\label{lem:ctx-trace-2}
For every pair of LPCF traces $(s, t)$ and LPCF programs $e_1, e_2 \in \prog(\tau)$, 
if $e_1 \transto t$ and $e_2 \transto t$, then for all linear context $\context_{x: \tau}$, 
$\context[e_1/x] \transto{(s, t)}$ implies $\context[e_2/x] \transto{(s, t)}$.
\end{lemma}
\ifLONG
\begin{proof}
We prove by induction on the full length of 
$\context[e_1 /x ] \transto{s}$, 
counting internal transitions. 

The base case is trivial. For non-empty traces, we analyze by cases:
\begin{itemize}
\item $\context[e_1/x] \reduce \context'[e_1 / x] \transto{(s, t)}$ with $\context \reduce \context'$.
  By induction $\context'[e_2/x] \transto{(s,t)}$,
  hence $\context[e_2/x] \reduce \context'[e_2/x] \transto{(s, t)}$.
\item $\context[e_1/x] \reduce \context[e_1' / x] \transto{(s, t)}$ with $e_1 \reduce e_1'$. 
  Clearly $e_1' \transto{t}$, so by induction, $\context[e_2 / x] \transto{(s, t)}$.
\item $\context[e_1/x] \intransto{\alpha} \context'[e_1'/y] \transto{(s, t')}$ with $t = \alpha \cdot t'$.
  Since $e_2 \transto t$, i.e., there exists $e_2'$ and $e_2''$ such that 
  $e_2 \reduce^* e_2'' \transto{\alpha} e_2' \transto{t'}$, by induction, we have 
  $\context'[e_2'/y] \transto{(s, t')}$.
  According to the definition of linear context reduction, $\context$ must be an evaluation context, 
  hence 
  $\context[e_2 / x] \reduce^* \context[e_2'' / x] \intransto{\alpha} \context'[e_2' / y] \transto{(s, t')}$, 
  i.e., $\context[e_2 / x] \transto{(s, t)}$. 
\item $\context[e_1/x] \transto{\alpha} \context'[e_1/x] \transto{(s', t)}$ 
  with $\context \lintransto\alpha \context'$ and $s = \alpha \cdot s'$. 
  By induction, $\context'[e_2/x] \transto{(s', t)}$, which  follows that 
  $\context[e_2/x]\transto{\alpha}\context'[e_2/x] \transto{(s', t)}$, 
  i.e., $\context[e_2 / x] \transto{(s, t)}$. 
\item $\context \equiv x$ and $s = t$. Clearly $\context[e_2/x] \equiv e_2 \transto t$, i.e., 
  $\context[e_2 / x] \transto{(s, t)}$. 
\end{itemize}
Lemma~\ref{lem:lcr} and Lemma~\ref{lem:lc-trans} ensure that the above analysis is comprehensive. 
\qed
\end{proof}
\fi

\ifLONG
\begin{proof}[Theorem~\ref{thm:precong-lpcf}]
\else 
\begin{theorem}[Linear precongruence of $\preord^T$]
\label{thm:precong-lpcf}
Trace preorder $\preord^T$ is a precongruence with respect to linear contexts, i.e., 
$e_1 \preord^T e_2$ implies that $\context[e_1 / x] \preord^T \context[e_2 / x]$ 
for all linear contexts $\context_x$.
\end{theorem}
\begin{proof}
\fi
Consider arbitrary linear context $\context$ and trace $s$ such that $\context[e_1/x] \transto s$. 
By Lemma~\ref{lem:ctx-trace}, $e_1$ has a $(\context, s)$-trace $t$, i.e., $e_1 \transto t$, 
which implies $e_2 \transto t$ since $e_1 \preord^T e_2$. 
By Lemma~\ref{lem:ctx-trace-2}, $\context[e_2/x] \transto {(s,t)}$, hence $\context[e_1/x] \preord^T \context[e_2/x]$.
\qed
\end{proof}

\begin{theorem}[Soundness of trace equivalence]\label{thm:tc}
In LPCF, it holds that $\simeq^T \;\subseteq\; \simeq^C$.
\end{theorem}
\begin{proof}
\ifLONG
For every well typed linear context $\context_x$, 
if $\context[e_1/x] \converge$, i.e. $\context[e_1/x] \reduce^* v$ for some canonical term $v$, 
then $\context[e_1/x] \reduce^* v \transto{\alpha}$ for some external action $\alpha$. 
By the precongruence property of $\preord^T$, Theorem~\ref{thm:precong-lpcf}, 
we have $\context[e_1/x] \preord^T \context[e_2/x]$. 
Therefore, there is some term $e$ such that $\context[e_2/x]\reduce^* e \transto{\alpha}$. 
In order to perform an external action, here $e$ must be a canonical term and it follows that 
$\context[e_2/x]\converge$.
Similarly we can show that if $\context[e_2/x] \converge$, then $\context[e_1/x] \converge$.
\else
It is easy to prove with the linear precongruence of trace preorder.
\fi
\qed
\end{proof}

\begin{theorem}[Completeness]\label{thm:ct}
In LPCF, it holds that $\simeq^C \;\subseteq\; \simeq^T$.
\end{theorem}
\ifLONG
\begin{proof}
We first notice that in Definition~\ref{def:preordC} the relations
$\preord^C$ and $\simeq^C$ are defined by quantifying over all linear
contexts. In fact, it suffices to quantify over the subset of
linear contexts that are evaluation contexts (viewing $\context_x$ as
$\context[[\ ]/x]$). In other words, for
any two terms of the same type,
\vskip 2mm
\begin{minipage}{11cm}
 (*) if they are distinguished by a linear
context, with
$\context[e_1/x]\converge$ but $\context[e_2/x]\diverge$, then they are
also distinguished by an evaluation context $\context'$ with
$\context\reduce^*\context'$. 
\end{minipage}
\vskip 2mm
\noindent This is proved as follows.
Suppose $\context[e_1/x]\converge$ but $\context[e_2/x]\diverge$.
We observe that all reduction
sequence starting from $\context$ must terminate in order to ensure
$\context[e_1/x]\converge$. So we can proceed by induction on the
length of the reduction sequence. 
\begin{itemize}
\item If $\context$ is already an evaluation context, then we are done by setting
$\context'$ to be $\context$.
\item $\context$ cannot be a normal product, a tensor product or an
  abstraction, as all these forms cannot be reduced any more, and are
  not able to meet the requirement that $e_2\diverge$.

\item For all other cases, if $\context\reduce\context_1$  then 
  $\context_1$ is also a linear context and by
  determinacy of reduction semantics, Proposition~\ref{prop:det}, we
  have $\context[e_1/x] \reduce \context_1[e_1/x] \converge$ and 
  $\context[e_2/x] \reduce \context_1[e_2/x] \diverge$. By induction
  applied to $\context_1$, there exists some evaluation context
  $\context'$ such that $\context_1\reduce^*\context'$, 
 $\context'[e_1/x] \converge$ and $\context'[e_2/x]\diverge$. 
Hence $\context\reduce^* \context'$ and we can find the required $\context'$.
\end{itemize}

We now show that, for any terms $e_1,e_2$ of the same type with $e_1 \simeq^C e_2$ and any action sequence $s$, 
if $e_1\transto{s}$ then $e_2\transto{s}$, which establishes $e_1 \preord^T e_2$. 
Similarly we can prove $e_2 \preord^T e_1$ but we shall omit the details.

We proceed by induction on the length of the transition $e_1 \transto{s}$. The base case is trivial. 
For the inductive step, we distinguish two cases.
\begin{itemize}
\item $e_1\reduce e'_1\transto{s}$. Clealy, we can prove, by induction on the structure of context, 
that $e'_1\preord^C e_1$, then $e'_1 \preord^C e_2$. By induction, we obtain that $e_2\transto{s}$.

\item $e_1\transto{\alpha}e'_1\transto{s}$. There are a few subcases, depending on the form of $\alpha$.
\begin{itemize}
\item $\alpha\equiv n$. 
  Both $e_1$ and $e_2$ have type $\tnat$, and $e_1 \transto{n} \infini$, so $e_1 \converge$ 
  and $e_1 \equiv n$.
  Because $e_1 \simeq^C e_2$, $e_2 \converge$ too (otherwise the simple linear context $x$ 
  can distinguish them). 
  We claim that for every possible reduction sequence $e_2 \reduce^* e_2' \not\reduce$, 
  $e_2' \equiv n$.
  First, because $e_2$ has type $\tnat$, by Proposition~\ref{prop:canonical}, $e_2'$ has to be an 
  integer constant. Assume that $e_2 \reduce^* m$ and $m \neq n$. Then the context 
  $$C_x \equiv \cif {x = n}  0 {\infini}$$ will distinguish $e_1$ from $e_2$, which contradicts 
  $e_1 \simeq^C e_2$.
  
  Similar is the case where $\alpha$ is a boolean constant.

\item $\alpha\equiv @e$. 
  In this case $e_1$ and $e_2$ must have a function type, and clearly 
  $e_1$ is in the canonical form: $e_1 \equiv \lambda x.e''_1$ and $e'_1 \equiv e''_1[e/x]$.
  Because $e_1 \simeq^C e_2$, the reduction of $e_2$ necessarily terminates and $e_2$ will be reduced to some canonical form 
$\lambda x.e''_2$, then $e_2 \transto{@e} e''_2[e/x]$. We claim that $e''_1[e/x] \preord^C e''_2[e/x]$.

Suppose for a contradiction that $e''_1[e/x] ~\not\preord^C~ e''_2[e/x]$. There exists some linear 
context $\context$ such that $\context[e''_1[e/x]/y]\Reduceto$ but $\context[e''_2[e/x]/y]\Uparrow$. 
By property (*) above, we can assume that $\context$ is  an evaluation context.
Then we can construct another context $\context':=\context[ye/y]$. 
Clearly $\context'[e_1/y]\Reduceto$ because 
\[
\context'[e_1/y]\equiv\context[e_1 e/y]\equiv\context[(\lambda x.e''_1)e/y]\reduce\context[e''_1[e/x]/y]\Reduceto.
\] 
However, $\context'[e_2/y]\Uparrow$ because 
\[
\context'[e_2/y]\equiv\context[e_2 e/y]\reduce^*\context[(\lambda x.e''_2)e/y]\reduce\context[e''_2[e/x]/y]\Uparrow.
\]
This is a contradiction to $e_1\preord^C e_2$. 
Therefore the assumption is wrong and we have $e''_1[e/x] ~\preord^C~ e''_2[e/x]$. 
By induction, we have $e''_2[e/x]\transto{s}$ and it follows that 
$e_2 \transto{\alpha}e''_2[e/x]\transto{s}$.

\item $\alpha\equiv \cproj_1$.  In this case $e_1,e_2$ must have a normal product type,
  then $e_1$ is in a canonical form $\cpair{ e_{11},e_{12}}$ and $e'_1\equiv e_{11}$. 
  The term $e_2$ can be reduced to a canonical term $\langle e_{21},e_{22}\rangle$, 
  and then $e_2\transto{\cproj_1}e_{21}$. We claim that $e_{11} \preord^C e_{21}$.

Suppose for a contradiction that $e_{11} ~\not\preord^C~ e_{21}$.
There exists some linear context $\context$ such that $\context[e_{11}/y]\Reduceto$ 
but $\context[e_{21}/y]\Uparrow$. By property (*), $\context$ can be
 assumed to be an evaluation context.
Then we can construct another context $\context':=\context[\cproj_1(y)/y]$. 
Clearly $\context'[e_1/y]\Reduceto$ because 
\[
\context'[e_1/y] \equiv \context[\cproj_1(e_1)/y]
\equiv \context[\cproj_1(\langle e_{11},e_{21}\rangle)/y] 
\reduce \context[e_{11}/y]\Reduceto.
\] 
However, $\context'[e_2/y]\Uparrow$ because 
\[
\context'[e_2/y] \equiv \context[\cproj_1(e_2)/y] \reduce^* 
\context[\cproj_1(\langle e_{21},e_{22}\rangle)/y] \reduce 
\context[e_{21}/y]\Uparrow.
\]
This is a contradiction to $e_1 \preord^C e_2$. Therefore the assumption is wrong 
and we have $e_{11} \preord^C e_{21}$. By induction, we have $e_{21}\transto{s}$ 
and it follows that $e_2\transto{\alpha}e_{21}\transto{s}$.

The case for $\alpha\equiv \cproj_2$ is similar.

\item $\alpha\equiv\otimes e$.  In this case $e_1,e_2$ must have a tensor product type, 
  then $e_1$ is in a canonical form $e_{11} \otimes  e_{12}$ and 
 $e'_1 \equiv e[e_{11}/x,e_{12}/y]$. The term $e_2$ can be reduced to
 a canonical term $e_{21}\otimes e_{22}$, and then $e_2\transto{\tensor e}e[e_{21}/x, e_{22}/y]$. We claim that
 $e[e_{11}/x,e_{12}/y] ~\preord^C~ e[e_{21}/x,e_{22}/y]$.

Suppose for a contradiction that $e[e_{11}/x,e_{12}/y] \not\preord^C e[e_{21}/x,e_{22}/y]$. 
There exists some linear context $\context$ such that $\context[e_1'/z] \Reduceto$ 
but $\context[(e[e_{21}/x,e_{22}/y])/z] \Uparrow$. By property (*) above, we can assume that $\context$ is  an evaluation context.
Then we can construct another context $\context':=\context[(\clet{x \otimes y}{z}{e})/z]$. 
Clearly $\context'[e_1/z]\Reduceto$ because 
\[
\begin{array}{ll}
\context'[e_1/z] 
& \equiv \context[(\clet{x \otimes y}{e_1}{e})/z] 
\\
& \equiv \context[(\clet{x \otimes y}{e_{11} \otimes e_{12}}{e})/z]
\\
& \reduce \context[(e[e_{11}/x,e_{12}/y])/z] \Reduceto.
\end{array}
\] 
However, $\context'[e_2/z]\Uparrow$ because 
\[
\begin{array}{ll}
\context'[e_2/z] 
& \equiv \context[(\clet{x \otimes y}{e_2}{e})/z]
\\
& \reduce^* \context[(\clet{x \otimes y}{e_{21}\otimes e_{22}}{e})/z]
\\
& \reduce \context[(e[e_{21}/x,e_{22}/y])/z]\Uparrow.
\end{array}
\]
This is a contradiction to $e_1\preord^C e_2$. Therefore the
assumption is wrong and we have $e[e_{11}/x,e_{12}/y] \preord^C e[e_{21}/x,e_{22}/y]$.
By induction, we have the transition $e[e_{21}/x,e_{22}/y]\transto{s}$. It follows
that $e_2\transto{\alpha} e[e_{21}/x,e_{22}/y]\transto{s}$.
\qed
 \end{itemize}
\end{itemize}
\end{proof}
\else
\begin{proof}
The proof is done by induction on the length of traces and by analysis of different forms 
of transitions --- the general way is to define a context that distinguishes the two terms 
when assuming that one of the term can take a particular trace while the other cannot.
The proof relies on the determinancy of program evaluation. 
The full proof is provided in~\cite{DZ11-full}.\qed 
\end{proof}
\fi

%% file: nlpcf.tex
\section{The non-deterministic linear PCF}
\label{sec:nlpcf}

In this section we shall extend our language with non-determinism, where emerges the example 
in Section~\ref{sec:intro}. We show that our approach can still be
applied to 
characterize linear contextual equivalence 
in the non-deterministic setting.  

The extension of non-determinism is made in Moggi's computational framework~\cite{Mog91}, 
which provides a call-by-value wrapping of imperative features in pure functional languages, 
using monadic types.
\ifLONG
We use Moggi's framework also because our original semantics of LPCF is a call-by-name 
evaluation strategy, while we need the call-by-value evaluation of non-deterministic choice 
for illustrating interesting effects. 
Were the original semantics call-by-value, we would not have to use Moggi's framework. 

\else
\fi
The types of the non-deterministic LPCF (NLPCF for short) are extended by a unary type constructor 
$\tcomp$ --- $\tcomp \tau$ is the type for non-deterministic computations that return, 
if terminate, values of type $\tau$. 
The language then has extra constructs related to non-determinism:
$$
\begin{array}{r@{\;}l@{\quad}l}
e, e', \ldots ::= & \ldots  & 
\\
\mid & \cval(e) & \mbox{Trivial computation}
\\
\mid & \cbind{x}{e}{e'} & \mbox{Sequential composition}
\\
\mid & e \choice e' & \mbox{Non-deterministic choice}
\end{array}
$$
$\cval(e)$ is the trivial computation that returns directly $e$ as a value; 
$\cbind{x}{e}{e'}$ binds the value of the (non-deterministic) computation $e$ to the variable $x$ 
and evaluates $e'$; 
$e \choice e'$ chooses non-deterministically a computation from $e$ and $e'$ and executes it. 

Type assertions for the extra constructs are defined by the following rules:
$$
\begin{array}{l}
 \begin{prooftree}
 \Gamma ; \Delta \vdash e : \tau 
 \justifies 
 \Gamma ; \Delta \vdash \cval(e) : \tcomp \tau 
 \end{prooftree}
 \qquad 
 \begin{prooftree}
 \Gamma ; \emptyset \vdash e_1 : \tcomp \tau_1  
 \quad 
 \Gamma, x: \tau_1 ; \Delta \vdash e_2 : \tcomp \tau_2
 \justifies 
 \Gamma ; \Delta \vdash \cbind{x}{e_1}{e_2} : \tcomp \tau_2
 \end{prooftree}
 \\[1.5em]
 \begin{prooftree}
 \Gamma ; \Delta \vdash e_1 : \tcomp \tau_1  
 \quad 
 \Gamma; \Delta' ,  x: \tau_1 \vdash e_2 : \tcomp \tau_2
 \justifies 
 \Gamma ; \Delta, \Delta' \vdash \cbind{x}{e_1}{e_2} : \tcomp \tau_2
 \end{prooftree}
 \qquad 
 \begin{prooftree}
 \Gamma ; \Delta \vdash e_i : \tcomp \tau \; (i = 1, 2)
 \justifies
 \Gamma ; \Delta \vdash e_1 \choice e_2 : \tcomp \tau
 \end{prooftree}
\end{array}
$$
The typing for sequential computation must respect the linearity restriction. 
Also, linear variables appear in both branches of the non-deterministic choice, since eventually 
only one branch will be executed. 

We write $\prog^{NL}(\tau)$ for the set of programs (closed terms) of type $\tau$ in NLPCF.

\subsection{Operational semantics}
The operational semantics of NLPCF is extended with the following basic reduction rules 
$$
\begin{array}{l}
\cbind{x}{\cval(e')}{e} \reduce (\lambda x \ldot e)e', \mbox{ where } e' \not\reduce ,
\\
e_1 \choice e_2 \reduce e_i \; (i = 1, 2) ,
\end{array}
$$
together with the extension for evaluation contexts:
$$
\evcontext ::= \ldots \mid \cbind{x}{\evcontext}{e} \mid \cval(\evcontext) .
$$
According to linearity, we do not allow evaluation contexts $\evcontext \choice e$ and 
$e \choice \evcontext$.

The $\choice$ operator behaves like the internal choice in CSP~\cite{Hoa85}. We can also add the external 
choice operator $\square$, together with rules
$$
\begin{array}{l}
e_1 \, \square \, e_2 \reduce e_1' \, \square \, e_2, \mbox{ where } e_1 \reduce e_1' ,
\\
e_1 \, \square \, e_2 \reduce e_1 \, \square \, e_2', \mbox{ where } e_2 \reduce e_2' .
\end{array}
$$
In accord with linearity, the typing rule for $\square$ will be different from that of $\choice$:
\[
\begin{prooftree}
  \Gamma ; \Delta_1 \vdash e_1 : \tcomp \tau 
  \quad 
  \Gamma ; \Delta_2 \vdash e_2 : \tcomp \tau 
  \justifies
  \Gamma ; \Delta_1, \Delta_2 \vdash e_1 \,\square\, e_2 : \tcomp \tau
\end{prooftree}
\]
Our later development only considers the internal choice operator, but it can be easily adapted 
to languages with the external choice, with careful treatment of the reduction which can 
discard linear variables.

Canonical terms of NLPCF, besides the canonical terms of LPCF, now include terms of the form 
$\cval(v)$ where $v \not\reduce$. 
The propositions about canonical form and subject reduction still hold.
\begin{proposition}
If $e$ is a NLPCF program and $e \not\reduce$, then $e$ must be in canonical form. 
\end{proposition}
\begin{proposition}
In NLPCF, if $\Gamma; \Delta \vdash e : \tau$ and $e \reduce e'$, 
then $\Gamma; \Delta \vdash e' : \tau$.
\end{proposition}

\ifLONG
The reduction system for NLPCF is non-deterministic and a term does not necessarily reduce
to a unique value even if it converges --- there is no confluence property in NLPCF.
For any closed term $e$, we say 
\begin{itemize}
\item $e$ \emph{may converge} (written as $e \converge$) 
if there exists a value $v$ such that  $e \reduce^* v \not\reduce$;
\item $e$ {\em must converge} (written as $e \allconverge$) 
\ifLONG
if there is no infinite reduction starting from $e$, i.e., a reduction of $e$ always terminates;  
\else
if there is no infinite reduction starting from $e$;
\fi
\item $e$ \emph{may diverge} (written as $e \diverge$) 
if $e$ has an infinite reduction sequence $e \reduce e_1 \reduce e_2 \reduce \cdots$;
\item $e$ {\em must diverge} (written as $e \alldiverge$) 
\ifLONG
if there is no value $v$ such that $e \reduce^* v \not\reduce$, i.e., $e$ never reduces to a value.
\else
if there is no value $v$ such that $e \reduce^* v \not\reduce$.
\fi
\end{itemize}
\else
The reduction system for NLPCF is non-deterministic and we say that a closed term $e$ {\em (may) converge}, 
still written as $e \converge$, if there exists a value $v$ such that $e \reduce^* v \not \reduce$.
\fi

\subsection{Labeled transition system}
The labeled transition system for NLPCF is extended by the following rule:
$$
\prooftree
\Gamma ; \Delta \vdash \cval(e) : \tcomp \tau 
\quad 
e \not\reduce
\justifies
\cval(e) \transto{\tcomp} e
\endprooftree
$$ 
The rule 
represents how programs of monadic types interact with contexts.

Similar as in LPCF, we can define trace, trace preorder (written as $\preord^{NT}$)
and trace equivalence (written as $\simeq^{NT}$) for NLPCF. 

\begin{example}
\label{example-1}
Consider the two programs $f_1$ and $f_2$ in Section~1. Both of them have, among many others, 
the trace $\cpair{\tcomp, @e, \tcomp, \mathtt 1}$ because of the following inference
\\
\parbox{15em}{
\begin{eqnarray*}
f_1 & \equiv & \cval(\lambda x.\cval(\mathtt 0)\choice\cval(\mathtt 1))\\
 & \transto{\tcomp} &  \lambda x.\cval(\mathtt 0)\choice\cval(\mathtt 1)\\
 & \transto{@ e} & (\cval(\mathtt 0)\choice\cval(\mathtt 1))[e/x]\\
 & \equiv & \cval(\mathtt 0)\choice\cval(\mathtt 1)\\
 & \reduce & \cval(\mathtt 1) \\
 & \transto{\tcomp} & \mathtt 1\\
 & \transto{\mathtt 1} & \pmb{\Omega} ,
\end{eqnarray*}
}
\qquad 
\parbox{15em}{
\begin{eqnarray*}
f_2 & \equiv & \cval(\lambda x.\cval(\mathtt 0))\choice\cval(\lambda x.\cval(\mathtt 1))\\
 & \reduce & \cval(\lambda x . \cval(\mathtt 1) )\\
 & \transto{\tcomp} &  \lambda x.\cval(\mathtt 1)\\
 & \transto{@ e} & \cval(\mathtt 1)[e/x]\\
 & \equiv & \cval(\mathtt 1)\\
 & \transto{\tcomp} & \mathtt 1\\
 & \transto{\mathtt 1} & \pmb{\Omega} .
\end{eqnarray*}
}
\end{example}

The definition of linear context is as in LPCF, so correspondingly we have the following linear 
context transitions:
\begin{eqnarray*}
\context[\cbind{z}{x}{e}/y] & \lintransto{\tcomp} & \context[(\lambda z . e)x'/y] 
\end{eqnarray*}
where $x'$ is a fresh variable.
The linear context transition lemma still holds:
\begin{lemma}[Linear context transition lemma in NLPCF]
For every linear context $\context_{x:\tau}$ and NLPCF program $e \in \prog^{NL}(\tau)$ 
such that $\context[e/x] \not\reduce$, 
a transition from $\context[e/x]$ must be either of the two forms:
\begin{itemize}
\item $\context[e/x] \transto{\alpha} \context'[e/x]$ with $\context \transto{\alpha} \context'$;
\item $\context \equiv x$ and $\context[e/x] \equiv e \transto{\alpha} e'$.
\end{itemize}
\end{lemma}
\ifLONG
\begin{proof}
Similar as in Lemma~\ref{lem:lc-trans}. 
\qed
\end{proof}
\fi

\subsection{Linear contextual equivalence in NLPCF}
\label{sec:ce-nlpcf}

\ifLONG
The Morris-style contextual equivalence depends on the notion of convergence,  
but in NLPCF, we need to choose between the may and must notions of convergence. 

The notions of convergence/divergence in NLPCF accordingly leads to the following notions of 
equivalence relations of programs. Let $e_1, e_2 \in \prog^{NL}(\tau)$ for arbitrary type $\tau$, 
\begin{itemize}
\item $e_1 \simeq^\converge e_2$:\quad $e_1 \converge$ if and only if $e_2 \converge$;
\item $e_1 \simeq^\allconverge e_2$:\quad $e_1 \allconverge$ if and only if $e_2 \allconverge$;
\item $e_1 \simeq^\diverge e_2$:\quad $e_1 \diverge$ if and only if $e_2 \diverge$;
\item $e_1 \simeq^{\alldiverge} e_2$:\quad $e_1 \alldiverge$ if and only if $e_2 \alldiverge$.
\end{itemize}
It can be easily checked that $\simeq^\converge \;=\; \simeq^{\alldiverge}$ and 
$\simeq^\diverge \;=\; \simeq^{\allconverge}$.

Must convergence equivalence $\simeq^\allconverge$ does not conform to trace equivalence 
in a non-deterministic language. 
If the reduction is deterministic or confluent, we can conclude that a term converges 
as long as it has non-empty traces, however it is not true for must convergence in 
a non-deterministic language --- by observing the traces of a term we can no longer tell 
whether a term has a non-terminating reduction sequence, since every term can take the empty trace, 
which by itself can represent divergence. 
In the contrary, if a term has {\em only} the empty trace, then we can conclude that the term 
must diverge. 

The linear contextual equivalence in NLPCF is defined based on the notion of may convergence.

\else
\fi
\begin{definition}[Non-deterministic linear contextual equivalence]
We write $e_1 \preord^{NC}_\tau e_2$ for $e_1, e_2 \in \prog^{NL}(\tau)$
if $\context[e_1 / x] \converge$ implies $\context[e_2 / x] \converge$ for all linear context 
$\context_{x:\tau}$. 
The relation $\preord^{NC}$ is called {\em non-deterministic linear contextual preorder}.
{\em Non-deterministic linear contextual equivalence $\simeq^{NC}$} is defined as the 
symmetrization of $\preord^{NC}$, that is, $e_1 \simeq^{NC}_\tau e_2$ iff $e_1 \preord^{NC}_\tau e_2$ 
and $e_2 \preord^{NC}_\tau e_1$.
\end{definition}
\ifLONG

\else
Note that the definition is based on the notion of convergence of NLPCF, which requires at least one terminating 
reduction sequence.  

\fi

The definition of linear context reductions remains the same as in LPCF, except that we are 
considering the extended transition system for NLPCF. 
The linear context reduction lemma still holds, from which the
precongruence of trace preorder follows, 
which in turn enables us to prove the soundness of trace preorder with
  respect to linear contextual equivalence in NLPCF.

\begin{lemma}[Linear context reduction lemma in NLPCF]
\label{lem:nlcr}
For every linear context $\context_{x:\tau}$ and NLPCF program $e \in \prog^{NL}(\tau)$, 
if $\context[e / x]$ is reducible, then $\context[e/x]\reduce$ must be a linear context reduction.
\end{lemma}
\ifLONG
\begin{proof}
The proof goes as in Lemma~\ref{lem:lcr}, by induction on the structure of linear context $\context$. 
We show only the cases for new constructs.
\begin{itemize}
\item $\context \equiv \cbind{y}{\context'}{e'}$.
  This is an evaluation context, so if $\context'[e/x]$ reduces, it must be a linear context reduction, 
  then $\context[e/x] \reduce$ is a linear context reduction of the same form as of $\context'[e/x] \reduce$.
  If $\context'[e/x]$ does not reduce, which must be canonical of the form $\cval(\cdots)$, 
  there are two cases:
  \begin{itemize}
  \item $\context' \equiv \cval(\context'')$ and $\context''[e/x] \not\reduce$. Then 
    \[
    \begin{array}{ll}
    \context[e/x] & \equiv \cbind{y}{\cval(\context''[e/x])}{e'} 
    \\
    & \reduce (\lambda y . e') \context''[e/x]
    \\
    & \equiv ((\lambda y . e') \context'') [ e / x] \quad 
    \mbox{(because $x$ does not appear freely in $\lambda y . e'$)}
    \end{array}
    \]
    with
    \[
    \context \equiv \cbind{y}{\cval(\context'')}{e'} 
    \reduce (\lambda y . e')\context'' ,
    \]
    which is a linear context.
    The reduction is the first form of LCR. 
  \item $\context' \equiv x$ and $e \equiv \cval(e'')$ ($e'' \not\reduce$). In this case, 
    $$
    \context[e/x] \equiv \cbind{y}{\cval(e'')}{e'} \reduce (\lambda y . e') e''.
    $$
    It is clear that $\context \equiv \cbind{y}{x}{e'} \lintransto{\tcomp} (\lambda y . e') x'$ 
    and $e \equiv \cval(e'') \transto{\tcomp} e''$, so the reduction is the third form 
    of LCR.
  \end{itemize}
\item $\context \equiv \cbind{y}{e'}{\context'}$.
  If $e' \reduce e''$, then $\context[e'/x] \reduce \cbind{y}{e''}{\context'[e/x]}$ with
  $\context \reduce \cbind{y}{e''}{\context'}$.
  If $e'$ does not reduce, it must be of the form $\cval(e'')$, then 
  $\context[e'/x] \reduce \context'[e/x][e''/y] \equiv \context'[e''/y][e/x]$, 
  with $\context \reduce \context'[e''/y]$, which is a linear context since $e''$ is closed.
  In both cases, the reduction is the first form of LCR.
\item $\context \equiv \context_1 \choice \context_2$. 
  Clearly, both $\context_1$ and $\context_2$ are linear contexts, then 
  $$
  \context[e/x] \equiv \context_1[e/x] \choice \context_2[e/x] \reduce \context_i[e/x] , \quad 
  (i = 1, 2), 
  $$
  with $\context \reduce \context_i$. The reduction is the first form of LCR.
\item $\context \equiv \cval(\context')$.
  Clearly $\context'$ is a linear context and $\context'[e/x] \reduce$ is a LCR, so 
  $\context[e/x] \reduce$ is also a LCR of the same form as $\context'[e/x] \reduce$.
\qed
\end{itemize}
\end{proof}
\fi

\begin{theorem}[Linear precongruence of $\preord^{NT}$]
\label{thm:precong-nlpcf}
Trace preorder $\preord^{NT}$ is a precongruence with respect to linear contexts, i.e., 
$e_1 \preord^{NT} e_2$ implies that $\context[e_1 / x] \preord^{NT} \context[e_2 / x]$ 
for all linear contexts $\context_x$ in NLPCF.
\end{theorem}

\begin{theorem}[Soundness of $\simeq^{NT}$]
In NLPCF, it holds that $\simeq^{NT} \;\subseteq\; \simeq^{NC}$.
\end{theorem}
\ifLONG
\begin{proof}
Assume that $e_1, e_2 \in \prog^{NL}(\tau)$ are two programs of NLPCF such that $e_1 \simeq^{NT} e_2$. 
By precongruence, for every linear context $\context_{x:\tau}$, 
$\context[e_1/x] \simeq^{NT} \context[e_2/x]$. If $\context[e_1/x] \converge$, 
i.e., $\trace(\context[e_1/x])$ has non-empty traces, then $\trace(\context[e_2/x])$ has non-empty 
traces too, hence $\context[e_2/x] \converge$. 
Similarly, if $\context[e_2/x] \converge$, then $\context[e_1/x] \converge$. 
\qed
\end{proof}
\fi

The above theorem allows us to prove the equivalence of the two functions in Example~1: 
it is easy to check that both functions have traces $\cpair{\tcomp, @e, \tcomp, \mathtt 0}$ 
and $\cpair{\tcomp, @e, \tcomp, \mathtt 1}$ (where $e$ is an arbitrary closed NLPCF term 
of proper type) as well as their subtraces, and they have no other traces.

\subsection{Completeness of trace equivalence in NLPCF}
\ifLONG
The rest of the section is devoted to proving the completeness of trace equivalence 
with respect to linear contextual equivalence in NLPCF.
\else
In NLPCF trace equivalence is still complete w.r.t. linear
contextual equivalence.
\fi
Unlike the proof of Theorem~\ref{thm:ct}, an induction over the length of traces does not 
work in a non-deterministic language, therefore we propose a novel proof for completeness.

We begin with constructing {\em trace-specific} linear contexts which ``recognize'' 
the corresponding traces. 
Given a trace $s$, we define the {\em $s$-context} $\context^s_{x: \tau}$ 
by induction on $s$:
\[\begin{array}{rcl}
\context^{\epsilon}_{x: \tau} & \defeq & 
{\cval(x)}
\\
\context^{\mathtt n}_{x:\tnat} & \defeq & 
\cif{x = \mathtt n}{\cval(\mathtt 0)}{\infini_{\tcomp \tnat}} 
\\
\context^{\ctrue}_{x: \tbool} & \defeq & 
\cif{x}{\cval(\mathtt  0)}{\infini_{\tcomp \tnat}} 
\\
\context^{\cfalse}_{x: \tbool} & \defeq & 
\cif{x}{\infini_{\tcomp \tnat}}{\cval(\mathtt  0)} 
\\
\context^{@e \cdot s}_{x: \tau \to \tau'} & \defeq & 
\cbind{y}{\cval(x e)}{\context^{s}_{y: \tau'}},
\mbox{ where $\emptyset; \emptyset \vdash e : \tau$} 
\\
\context^{\cproj_i \cdot s}_{x: \tau_1 \product \tau_2} & \defeq & 
\cbind{y}{\cval(\cproj_i(x))}{\context^s_{y: \tau_i}} 
\\
\context^{\otimes e \cdot s}_{x: \tau_1 \tensor \tau_2} 
& \defeq & \cbind{y}{\cval(\clet{z_1 \otimes z_2}{x}{e})}{\context^s_{y: \tau'}}, 
\\ & & \quad 
\mbox{ where $\emptyset; z_1: \tau_1, z_2: \tau_2 \vdash e : \tau'$}\\
\context^{\tcomp \cdot s}_{x: \tcomp \tau} & \defeq & 
\cbind{y}{x}{\context^s_{y: \tau}} 
\end{array}\]
It can be easily checked that 
 $\emptyset; x:\tau \vdash \context^s_{x: \tau}: \tcomp\tau'$ 
for some type $\tau'$, if $x$ is a linear variable,
and we call it a linear $s$-context.
In particular, if $s$ is a computational trace then $\tau'$ is $\tnat$.
We shall often omit the type information when it is obvious or irrelevant.

In the definition we do not consider traces $c \cdot s$ with boolean/integer constant $c$ 
followed by non-empty trace $s$, because a valid trace must be taken by a program, 
while a program that takes the $c$-transition must be $c$ itself, which no longer takes 
any external action after the transition ($c \transto{c} \infini$).

\ifLONG
The following two lemmas show that a program can take a computational trace $s$ if and only if the 
corresponding linear $s$-context, when filled in with the program, {\em may converge}.   

\begin{lemma}\label{lem:tr.cov}
For every NLPCF program $e$ and every computational trace $s$, if $e \transto{s}$ then $\context^s_x[e/x] \converge$, 
for linear $s$-context $\context^s_x$.
\end{lemma}
\begin{proof}
Let $e \in \prog^{NL}(\tau)$ be an arbitrary NLPCF program.
We prove by induction on the length of $s$.
\begin{itemize}

\item $s =  c$, where $c$ is a boolean or integer constant. We show the case of integer constant; 
  the proof for the boolean constant is similar. 
  If $e$ has the trace $c \cdot s'$, i.e., $e\reduce^* {c}$ and $s' = \nulltrace$, it follows that 
\[
\begin{array}{rcl}
\context^s_x[e/x] & \equiv & \cif{e = c}{\cval(\mathtt 0)}{\infini} \\
& \reduce^* & \cif{c = c}{\cval(\mathtt 0)}{\infini} \\
& \reduce^* & {\cval(\mathtt 0)} \converge .
\end{array}
\]

\item $s \equiv  @e'\cdot s'$. If $e$ has the trace $@e' \cdot s'$, i.e.,  
$$
e\reduce^* \lambda z.e_1 \xrightarrow{\,@e'\,} e_1[e'/z] 
\reduce^* e'' \xrightarrow{\,s'\,} , 
$$
with $e'$ a closed term of proper type and $e'' \not\reduce$, 
then it follows that 
\[
\begin{array}{l@{\;}l}
\context^s_x[e/x] & \equiv \cbind{y}{\cval(e\ e')}{\context^{s'}_y}\\
& \reduce^*  \cbind{y}{\cval((\lambda z.e_1) e')}{\context^{s'}_y}\\
& \reduce  \cbind{y}{\cval(e_1[e'/z])}{\context^{s'}_y}\\
& \reduce^* \cbind{y}{\cval(e'')}{\context^{s'}_y} \\
& \reduce^* \context^{s'}_y[e''/y]
\end{array}
\]
Since $s'$ is a shorter trace than $s$, by induction, we know from $e''\transto{s'}$ 
that $\context^{s'}_y[e''/y] \converge$, therefore $\context^s_x[e/x] \converge$.

\item $s =  \cproj_1 \cdot s'$. If $e$ has the trace $\cproj_1 \cdot s'$, i.e., 
$$
e\reduce^* \cpair{e_1,e_2} \xrightarrow{\,\cproj_1\,} e_1
\reduce^* e_1' \xrightarrow{\,s'\,} , 
$$
with $e_1' \not\reduce$, then it follows that 
\[\begin{array}{rcl}
\context^s_x[e/x] & \equiv & \cbind{y}{\cval(\cproj_1(e))}{\context^{s'}_y}\\
& \reduce^* & \cbind{y}{\cval(\cproj_1(\cpair{e_1,e_2}))}{\context^{s'}_y}\\
& \reduce & \cbind{y}{\cval(e_1)}{\context^{s'}_y}\\
& \reduce^* & \cbind{y}{\cval(e_1')}{\context^{s'}_y}\\
& \reduce^* & \context^{s'}_y[e_1'/y]
\end{array}\]
Since $s'$ is a shorter trace than $s$, by induction, we know from $e_1' \xrightarrow{\,s'\,}$ 
that $\context^{s'}_y[e_1'/y] \converge$, therefore $\context^s_x[e/x] \converge$.

The case $s = \cproj_2\cdot s'$ is similar.

\item $s = \otimes e' \cdot s'$. If $e$ has the trace $\otimes e'\cdot s'$, i.e., 
\begin{equation}\label{eq:ct3}
e\reduce^* e_1\otimes e_2 \xrightarrow{\,\otimes e'\,} e'[e_1/{z_1},e_2/{z_2}]
\reduce^* e'' \xrightarrow{\,s'\,},
\end{equation}
with $\emptyset; z_1: \tau_1, z_2: \tau_2 \vdash e': \tau'$ ($\tau \equiv \tau_1 \tensor \tau_2$) 
and $e'' \not\reduce$, then it follows that 
\[\begin{array}{rcl}
\context^s_x[e/x] & \equiv & \cbind{y}{\cval(\clet{z_1\otimes z_2}{e}{e'})}{\context^{s'}_y}\\
& \reduce^* & \cbind{y}{\cval(\clet{z_1\otimes z_2}{e_1\otimes e_2}{e'})}{\context^{s'}_y}\\
& \reduce & \cbind{y}{\cval(e'[e_1/{z_1},e_2/{z_2}])}{\context^{s'}_y}\\
& \reduce^* & \cbind{y}{\cval(e'')}{\context^{s'}_y} \\
& \reduce^* & \context^{s'}_y[e''/y]
\end{array}\]
Since $s'$ is a shorter trace than $s$, by induction, we know from 
$e'' \xrightarrow{\,s'\,}$ that $\context^{s'}_y[e''/y] \converge$, therefore $\context^s_x[e/x] \converge$.

\item $s =  \tcomp \cdot s'$. If $e$ has the trace $\tcomp \cdot s'$, i.e., 
$$
e\reduce^* \cval(e') \xrightarrow{\,\tcomp\,} e'
\xrightarrow{\,s'\,} ,
$$
with $e' \not\reduce$, then it follows that 
\[
\begin{array}{l@{\;}l}
\context^s_x[e/x] & \equiv \cbind{y}{e}{\context^{s'}_y}\\
& \reduce^* \cbind{y}{\cval(e')}{\context^{s'}_y}\\
& \reduce^* \context^{s'}_y[e'/y]
\end{array}
\]
Since $s'$ is a shorter trace than $s$, by induction, we know from $e' \xrightarrow{\,s'\,}$ 
that $\context^{s'}_y[e'/y] \converge$, therefore $\context^s_x[e/x] \converge$.
\qed
\end{itemize}
\end{proof}

\begin{lemma}\label{lem:cov.trace}
For any $e\in\prog^{NL}(\tau)$ and trace $s$, if $\context^s_x[e/x] \converge$ then $e \transto{s}$.
\end{lemma}
\begin{proof}
We prove by induction over the length of $s$, with an NLPCF program $e$.
\begin{itemize}
\item $s = \nulltrace$. It is clear that $\nulltrace \in \trace(e)$.

\item $s = c$, where $c$ is a boolean or integer constant. Assume that $c$ is an integer (the case of 
  booleans is similar). 
  Since $\context^s_x [e/x] \equiv \cif{e = c} {\cval(\mathtt 0)} {\infini} \converge$, it must hold 
  that $e$ may converge and $e \reduce^* c$, hence $e \reduce^* c \transto{c}$.

\item $s = @e' \cdot s'$, with $e'$ a closed term of proper type. 
  Since
  $$\context^{@e' \cdot s'}_x [e/x] \equiv \cbind{y}{\cval(e \ e')}{\context_y^{s'}} \converge ,$$ 
  there must be a reduction sequence
  \[
  \begin{array}{l@{\;}l}
  \context^{@e' \cdot s'}_x [e/x] 
  & \reduce^* \cbind{y}{\cval((\lambda z . e_1) e')}{\context_y^{s'}} 
  \mbox{ (where $e \reduce^* \lambda z . e_1$ )}  
  \\
  & \reduce \cbind{y}{\cval(e_1[e'/z])}{\context_y^{s'}} 
  \\
  & \reduce^* \cbind{y}{\cval(e'')}{\context_y^{s'}} 
  \mbox{ (where $e_1[e'/z] \reduce^* e''$ and $e'' \not\reduce$)}  
  \\
  & \reduce^* \context_y^{s'}[e''/y]
  \end{array}
  \]
  and $\context^{s'}_y[e''/y] \converge$, which implies that $e'' \transto{s'}$ by induction.
  Clearly, $e$ may converge and $e \reduce^* \lambda z . e_1 \transto{@e'} e_1[e'/z] \reduce^* e'' \transto{s'}$, 
  i.e., $e \transto{s}$. 

\item $s = \cproj_1 \cdot s'$. Since
  $$\context^{\cproj_1 \cdot s'}_x [e/x] \equiv \cbind{y}{\cval(\cproj_1(e))}{\context_y^{s'}} \converge ,$$ 
  there must be a reduction sequence
  \[
  \begin{array}{l@{\;}l}
  \context^{\cproj_1 \cdot s'}_x [e/x] 
  & \reduce^* \cbind{y}{\cval(\cproj_1(\cpair{e_1, e_2})) }{\context_y^{s'}} 
  \mbox{ (where $e \reduce^* \cpair{e_1, e_2}$ )}  
  \\
  & \reduce \cbind{y}{\cval(e_1)}{\context_y^{s'}} 
  \\
  & \reduce^* \cbind{y}{\cval(e_1')}{\context_y^{s'}} 
  \mbox{ (where $e_1 \reduce^* e_1'$ and $e_1' \not\reduce$)}  
  \\
  & \reduce^* \context_y^{s'}[e_1'/y]
  \end{array}
  \]
  and $\context_y^{s'}[e_1'/y] \converge$, which implies that $e_1' \transto{s'}$ by induction.
  Clearly, $e$ may converge and $e \reduce^* \cpair{e_1, e_2} \transto{\cproj_1} e_1 \reduce^* e_1' \transto{s'}$, 
  i.e., $e \transto{s}$. 

  The case $s \equiv  \cproj_2 \cdot s'$ is similar.

\item $s = \tensor e' \cdot s'$. Since
  $$
  \context^{\tensor e' \cdot s'}_x [e/x] \equiv 
  \cbind{y}{\cval(\clet{z_1 \tensor z_2}{e}{e'})}{\context_y^{s'}} \converge ,
  $$ 
  there must be a reduction sequence
  \[
  \begin{array}{l@{\;}l}
  \context^{\tensor e' \cdot s'}_x [e/x] 
  & \reduce^* \cbind{y}{\cval(\clet{z_1 \tensor z_2}{e_1 \tensor e_2}{e'})}{\context_y^{s'}} 
  \\ 
  & \qquad 
  \mbox{ (where $e \reduce^* e_1 \tensor e_2$ )}  
  \\
  & \reduce \cbind{y}{\cval(e'[e_1/z_1, e_2/z_2])}{\context_y^{s'}} 
  \\
  & \reduce^* \cbind{y}{\cval(e'')}{\context_y^{s'}} 
  \\
  & \qquad 
  \mbox{ (where $e'[e_1/z_1, e_2/z_2] \reduce^* e''$ and $e'' \not\reduce$)}  
  \\
  & \reduce^* \context_y^{s'}[e''/y]
  \end{array}
  \]
  and $\context_y^{s'}[e''/y] \converge$, which implies that $e'' \transto{s'}$ by induction.
  Clearly, $e$ may converge and 
  $e \reduce^* e_1 \tensor e_2 \transto{\tensor e'} 
    e'[e_1/z_1, e_2/z_2] \reduce^* e'' \transto{s'}$, 
  i.e., $e \transto{s}$. 

\item $s = \tcomp \cdot s'$.   Since
  $$\context^{\tcomp \cdot s'}_x [e/x] \equiv \cbind{y}{e}{\context_y^{s'}} \converge ,$$ 
  there must be a reduction sequence
  \[
  \begin{array}{l@{\;}l}
  \context^{\tcomp' \cdot s'}_x [e/x] 
  & \reduce^* \cbind{y}{\cval(e')}{\context_y^{s'}} 
  \quad \mbox{(where $e \reduce^* \cval(e')$ and $e' \not\reduce$)}  
  \\
  & \reduce^* \context_y^{s'}[e'/y]
  \end{array}
  \]
  and $\context^{s'}_y[e'/y] \converge$, which implies that $e' \transto{s'}$ by induction.
  Clearly, $e$ may converge and $e \reduce^* \cval(e') \transto{\tcomp} e' \transto{s'}$, 
  i.e., $e \transto{s}$. 
  \qed
\end{itemize}
\end{proof}

The next two lemmas act as the counterparts of the previous two, but our focus now is on traces that are not computational.
\begin{lemma}\label{lem:non-cmptr1}
If an NLPCF program $e$ has the trace $s\cdot\alpha$ with $e \xrightarrow{\,s\,} e' \xrightarrow{\,\alpha\,}$ and $e'\not\reduce$, then $\context^s_x[e/x] \reduce^* \cval(e')$.
\end{lemma}
\begin{proof}
We first note that $s$ is not a computational trace. Otherwise the program $e'$ derived from a computational trace would be $\Omega$, which cannot make an external action $\alpha$, a contradiction to the hypothesis that $e' \xrightarrow{\,\alpha\,}$.
 
Let $e \in \prog^{NL}(\tau)$ be an arbitrary NLPCF program.
Similar to the proof of Lemma~\ref{lem:tr.cov}, we prove by induction on the length of $s$.
\begin{itemize}
\item $s \equiv \nulltrace$. Clearly, it always holds that $e \xrightarrow{\,\nulltrace\,} e$ and
 $\context^\epsilon_x[e/x] \equiv \cval(\mathtt e) \reduce^*\cval(e)$.

\item $s \equiv  @e_1\cdot s'$. If $e$ has the trace $@e_1 \cdot s'$, i.e.,  
$$
e\reduce^* \lambda z.e_2 \xrightarrow{\,@e_1\,} e_2[e_1/z] 
\reduce^* e'' \xrightarrow{\,s'\,} e', 
$$
with $e_1$ a closed term of proper type and $e'' \not\reduce$, 
then it follows that 
\[
\begin{array}{l@{\;}l}
\context^s_x[e/x] & \equiv \cbind{y}{\cval(e\ e_1)}{\context^{s'}_y}\\
& \reduce^*  \cbind{y}{\cval((\lambda z.e_2) e_1)}{\context^{s'}_y}\\
& \reduce  \cbind{y}{\cval(e_2[e_1/z])}{\context^{s'}_y}\\
& \reduce^* \cbind{y}{\cval(e'')}{\context^{s'}_y} \\
& \reduce^* \context^{s'}_y[e''/y]
\end{array}
\]
Since $s'$ is a shorter trace than $s$, by induction, we know from $e''\transto{s'} e' \transto{\alpha}$ 
that $\context^{s'}_y[e''/y] \reduce^* \cval(e')$, therefore $\context^s_x[e/x] \reduce^* \cval(e')$ by transitivity of the relation $\reduce^*$.

\item $s =  \cproj_1 \cdot s'$. If $e$ has the trace $\cproj_1 \cdot s'$, i.e., 
$$
e\reduce^* \cpair{e_1,e_2} \xrightarrow{\,\cproj_1\,} e_1
\reduce^* e_1' \xrightarrow{\,s'\,} e', 
$$
with $e_1' \not\reduce$, then it follows that 
\[\begin{array}{rcl}
\context^s_x[e/x] & \equiv & \cbind{y}{\cval(\cproj_1(e))}{\context^{s'}_y}\\
& \reduce^* & \cbind{y}{\cval(\cproj_1(\cpair{e_1,e_2}))}{\context^{s'}_y}\\
& \reduce & \cbind{y}{\cval(e_1)}{\context^{s'}_y}\\
& \reduce^* & \cbind{y}{\cval(e_1')}{\context^{s'}_y}\\
& \reduce^* & \context^{s'}_y[e_1'/y]
\end{array}\]
Since $s'$ is a shorter trace than $s$, by induction, we know from $e_1' \xrightarrow{\,s'\,} e'\transto{\alpha}$ 
that $\context^{s'}_y[e_1'/y] \reduce^* \cval(e')$, therefore $\context^s_x[e/x] \reduce^* \cval(e')$ by transitivity of $\reduce^*$.

The case $s = \cproj_2\cdot s'$ is similar.

\item $s = \otimes e'' \cdot s'$. If $e$ has the trace $\otimes e''\cdot s'$, i.e., 
\begin{equation}\label{eq:ct33}
e\reduce^* e_1\otimes e_2 \xrightarrow{\,\otimes e''\,} e''[e_1/{z_1},e_2/{z_2}]
\reduce^* e''' \xrightarrow{\,s'\,}e',
\end{equation}
with $\emptyset; z_1: \tau_1, z_2: \tau_2 \vdash e'': \tau'$ ($\tau \equiv \tau_1 \tensor \tau_2$) 
and $e''' \not\reduce$, then it follows that 
\[\begin{array}{rcl}
\context^s_x[e/x] & \equiv & \cbind{y}{\cval(\clet{z_1\otimes z_2}{e}{e''})}{\context^{s'}_y}\\
& \reduce^* & \cbind{y}{\cval(\clet{z_1\otimes z_2}{e_1\otimes e_2}{e''})}{\context^{s'}_y}\\
& \reduce & \cbind{y}{\cval(e''[e_1/{z_1},e_2/{z_2}])}{\context^{s'}_y}\\
& \reduce^* & \cbind{y}{\cval(e''')}{\context^{s'}_y} \\
& \reduce^* & \context^{s'}_y[e'''/y]
\end{array}\]
Since $s'$ is a shorter trace than $s$, by induction, we know from 
$e''' \xrightarrow{\,s'\,}e'\transto{\alpha}$ that $\context^{s'}_y[e'''/y] \reduce^* \cval(e')$, therefore $\context^s_x[e/x] \reduce^* \cval(e')$.

\item $s =  \tcomp \cdot s'$. If $e$ has the trace $\tcomp \cdot s'$, i.e., 
$$
e\reduce^* \cval(e'') \xrightarrow{\,\tcomp\,} e''
\xrightarrow{\,s'\,} e',
$$
with $e' \not\reduce$, then it follows that 
\[
\begin{array}{l@{\;}l}
\context^s_x[e/x] & \equiv \cbind{y}{e}{\context^{s'}_y}\\
& \reduce^* \cbind{y}{\cval(e'')}{\context^{s'}_y}\\
& \reduce^* \context^{s'}_y[e''/y]
\end{array}
\]
Since $s'$ is a shorter trace than $s$, by induction, we know from $e'' \xrightarrow{\,s'\,}e'\transto{\alpha}$ 
that $\context^{s'}_y[e'/y] \reduce^* \cval(e')$, therefore $\context^s_x[e/x] \reduce^* \cval(e')$.
\qed
\end{itemize}
\end{proof}

\begin{lemma}\label{lem:non-cmptr2}
For every NLPCF program $e \in \prog^{NL}(\tau)$ and trace $s$ that is not computational, if $\context_{x}^s[e/x] \converge$ then there is some program $e'$ such that $e \xrightarrow{\,s\,} e'$ and $e' \converge$.
\end{lemma}
\begin{proof}
Similar to the proof of Lemma~\ref{lem:cov.trace}.
We prove by induction over the length of $s$, with an NLPCF program $e$.
\begin{itemize}
\item $s = \nulltrace$. Then $\context^s_x[e/x]\equiv\cval(e)\converge$. It means that $e\converge$. Clearly we also have $e\transto{\nulltrace} e$.

\item $s = @e'' \cdot s'$, with $e''$ a closed term of proper type. 
  Since
  $$\context^{@e'' \cdot s'}_x [e/x] \equiv \cbind{y}{\cval(e \ e'')}{\context_y^{s'}} \converge ,$$ 
  there must be a reduction sequence
  \[
  \begin{array}{l@{\;}l}
  \context^{@e'' \cdot s'}_x [e/x] 
  & \reduce^* \cbind{y}{\cval((\lambda z . e_1) e'')}{\context_y^{s'}} 
  \mbox{ (where $e \reduce^* \lambda z . e_1$ )}  
  \\
  & \reduce \cbind{y}{\cval(e_1[e''/z])}{\context_y^{s'}} 
  \\
  & \reduce^* \cbind{y}{\cval(e''')}{\context_y^{s'}} 
  \mbox{ (where $e_1[e''/z] \reduce^* e'''$ and $e''' \not\reduce$)}  
  \\
  & \reduce^* \context_y^{s'}[e'''/y]
  \end{array}
  \]
  and $\context^{s'}_y[e'''/y] \converge$, which implies that $e''' \transto{s'} e'$ and $e'\converge$ by induction.
 Therefore, $e \reduce^* \lambda z . e_1 \transto{@e'} e_1[e'/z] \reduce^* e'' \transto{s'} e'$, 
  i.e., $e \transto{s} e'$. 

\item $s = \cproj_1 \cdot s'$. Since
  $$\context^{\cproj_1 \cdot s'}_x [e/x] \equiv \cbind{y}{\cval(\cproj_1(e))}{\context_y^{s'}} \converge ,$$ 
  there must be a reduction sequence
  \[
  \begin{array}{l@{\;}l}
  \context^{\cproj_1 \cdot s'}_x [e/x] 
  & \reduce^* \cbind{y}{\cval(\cproj_1(\cpair{e_1, e_2})) }{\context_y^{s'}} 
  \mbox{ (where $e \reduce^* \cpair{e_1, e_2}$ )}  
  \\
  & \reduce \cbind{y}{\cval(e_1)}{\context_y^{s'}} 
  \\
  & \reduce^* \cbind{y}{\cval(e_1')}{\context_y^{s'}} 
  \mbox{ (where $e_1 \reduce^* e_1'$ and $e_1' \not\reduce$)}  
  \\
  & \reduce^* \context_y^{s'}[e_1'/y]
  \end{array}
  \]
  and $\context_y^{s'}[e_1'/y] \converge$, which implies that $e_1' \transto{s'} e'$ and $e'\converge$ by induction.
  Therefore,  $e \reduce^* \cpair{e_1, e_2} \transto{\cproj_1} e_1 \reduce^* e_1' \transto{s'} e'$, 
  i.e., $e \transto{s} e'$. 

  The case $s \equiv  \cproj_2 \cdot s'$ is similar.

\item $s = \tensor e'' \cdot s'$. Since
  $$
  \context^{\tensor e'' \cdot s'}_x [e/x] \equiv 
  \cbind{y}{\cval(\clet{z_1 \tensor z_2}{e}{e''})}{\context_y^{s'}} \converge ,
  $$ 
  there must be a reduction sequence
  \[
  \begin{array}{l@{\;}l}
  \context^{\tensor e'' \cdot s'}_x [e/x] 
  & \reduce^* \cbind{y}{\cval(\clet{z_1 \tensor z_2}{e_1 \tensor e_2}{e''})}{\context_y^{s'}} 
  \\ 
  & \qquad 
  \mbox{ (where $e \reduce^* e_1 \tensor e_2$ )}  
  \\
  & \reduce \cbind{y}{\cval(e''[e_1/z_1, e_2/z_2])}{\context_y^{s'}} 
  \\
  & \reduce^* \cbind{y}{\cval(e''')}{\context_y^{s'}} 
  \\
  & \qquad 
  \mbox{ (where $e''[e_1/z_1, e_2/z_2] \reduce^* e'''$ and $e''' \not\reduce$)}  
  \\
  & \reduce^* \context_y^{s'}[e'''/y]
  \end{array}
  \]
  and $\context_y^{s'}[e'''/y] \converge$, which implies that $e''' \transto{s'} e'$ and $e'\converge$ by induction.
Therefore, 
  $e \reduce^* e_1 \tensor e_2 \transto{\tensor e''} 
    e''[e_1/z_1, e_2/z_2] \reduce^* e''' \transto{s'}e'$, 
  i.e., $e \transto{s} e'$. 

\item $s = \tcomp \cdot s'$.   Since
  $$\context^{\tcomp \cdot s'}_x [e/x] \equiv \cbind{y}{e}{\context_y^{s'}} \converge ,$$ 
  there must be a reduction sequence
  \[
  \begin{array}{l@{\;}l}
  \context^{\tcomp' \cdot s'}_x [e/x] 
  & \reduce^* \cbind{y}{\cval(e'')}{\context_y^{s'}} 
  \quad \mbox{(where $e \reduce^* \cval(e'')$ and $e'' \not\reduce$)}  
  \\
  & \reduce^* \context_y^{s'}[e''/y]
  \end{array}
  \]
  and $\context^{s'}_y[e''/y] \converge$, which implies that $e'' \transto{s'} e'$ and $e'\converge$ by induction.
Therefore,  $e \reduce^* \cval(e'') \transto{\tcomp} e'' \transto{s'} e'$, 
  i.e., $e \transto{s} e'$. 
  \qed
\end{itemize}
\end{proof}
\else 
We can then show several facts about the convergence of $\context^{s}_x[e/x]$ depending on $s$, where $e$ must 
take the trace $s$. These facts consequently support our completeness proof. 
\fi

\begin{theorem}[Completeness of $\simeq^{NT}$]
In NLPCF, it holds that $\simeq^{NC} \;\subseteq\; \simeq^{NT}$.
\end{theorem}
\begin{proof}
\ifLONG
Assume that $e_1,e_2$ are two programs and $e_1 \simeq^{NC}
e_2$. Suppose $e_1 \xrightarrow{\,s\,}$ for some trace $s$. 
We distinguish two cases.
\begin{itemize}
\item $s$ is a computational trace.
By
Lemma~\ref{lem:tr.cov}, we have $\context^s_x[e_1/x] \converge$. Since
$e_1 \simeq^{NC} e_2$, it must be the case that $\context^s_x[e_2/x]
\converge$. By Lemma~\ref{lem:cov.trace}, it follows that $e_2
\xrightarrow{\,s\,}$.  

\item $s$ is not a computational trace. 
If $s=\nulltrace$, we obviously have $e_2\transto{s}$. Now suppose that $s=s'\cdot\alpha$, that is $e_1\transto{s'} e'_1\transto{\alpha}$ for some $e'_1\not\reduce$. By Lemma~\ref{lem:non-cmptr1} we have $\context^{s'}_x[e_1/x]\reduce^*\cval(e')$, which means that $\context^{s'}_x[e_1/x]\converge$. Since $e_1 \simeq^{NC}
e_2$, it must be the case that $\context^{s'}_x[e_2/x]\converge$. By Lemma~\ref{lem:non-cmptr2}, there is some $e'_2$ such that $e_2\transto{s'}e'_2$ and $e'_2\converge$. By Lemma~\ref{lem:tr.type}, which also holds for NLPCF, we see that $e'_2$ has the same type as $e'_1$. Since $s$ is not a computational trace, $\alpha$ must be in one of the forms $@e$, $\cproj_i$, $\otimes e$ or $\tcomp$. Depending on the type of $e_1$, in each case there exists some $e''_2$ such that $e''_2\not\reduce$ and $e'_2\reduce^* e''_2\transto{\alpha}$. It follows that $e_2\transto{s'}e'_2\reduce^*e''_2\transto{\alpha}$, that is $e_2\transto{s}$.
\end{itemize}

Symmetrically, any trace of $e_2$ is also a
trace of $e_1$. Therefore, we obtain $e_1 \simeq^{NT}
e_2$.

\else
Assume that $e_1,e_2$ are two programs and $e_1 \simeq^{NC} e_2$. Suppose $e_1 \xrightarrow{\,s\,}$ 
for some trace $s$. The proof distinguishes two cases: $s$ is either computational or non-computational. 
In both cases we can show that $e_2 \transto{s}$. See more details in~\cite{DZ11-full}.
\fi
\qed
\end{proof}

%% file: conclu.tex
\section{Conclusion}
\label{sec:conclu}
\ifLONG
We have presented a novel approach for characterizing program equivalence in linear contexts, 
via trace equivalence in appropriate labeled transition systems. The technique is both sound and 
complete, and as we have shown in the paper, is general enough to be adapted for languages with 
linear type systems. 
\else
\fi

Linear contextual equivalence is indeed a restricted notion of program equivalence and one may 
question its use in practice. As we have explained in the beginning of the paper, it does have 
application in security since we can use linearity to limit adversaries' behaviour. 
We also believe that such a notion of program equivalence can be useful in reasoning 
about programs in systems where only restricted access to resources is allowed, particularly when 
side effects are present. 
The result in non-deterministic languages already enables us to prove linear contextual equivalence 
between non-trivial programs. 

We have used both program transitions and context transitions to model the interactions between 
programs and contexts, and the program/context traces (if combined in a proper way) resembles 
strategies in game semantics~\cite{AM98,HO00}, despite of our operational treatment of traces. 
However, it is unclear whether the correspondence can be made between program/context actions 
in the trace model and player/oppenent moves in the game model --- the exact connection remains 
to clarify.